%% file: wmac_version.tex
\documentclass[letterpaper]{article} 
\usepackage{aaai2026}  
\usepackage{times}  
\usepackage{helvet}  
\usepackage{courier}  
\usepackage[hyphens]{url}  
\usepackage{graphicx} 
\usepackage{natbib}  
\usepackage{caption} 
\usepackage{algorithm}
\usepackage{algorithmic}

\usepackage{subfig}
\usepackage{amsmath, amsfonts}
\usepackage{booktabs}
\usepackage{multirow}
\usepackage[table]{xcolor}
\usepackage{xcolor}
\definecolor{lightpink}{RGB}{255, 210, 210}
\usepackage{newfloat}
\usepackage{listings}
\DeclareCaptionStyle{ruled}{labelfont=normalfont,labelsep=colon,strut=off} 
\floatstyle{ruled}
\newfloat{listing}{tb}{lst}{}
\floatname{listing}{Listing}
\nocopyright 

\usepackage{fancyhdr}
\fancypagestyle{firstpagehf}
{
   \fancyhf{}
   \fancyhead[HC]{The AAAI 2026 Bridge Program on Advancing Large Language Models and Multi-Agent Systems}
   \fancyfoot[C]{\thepage}
}

\title{AED: \underline{A}utomatic Discovery of \underline{E}ffective and \underline{D}iverse Vulnerabilities for Autonomous Driving Policy with Large Language Models}
\author {
    Le Qiu\textsuperscript{\rm 1}$^*$,
    Zelai Xu\textsuperscript{\rm 1}$^*$,
    Qixin Tan\textsuperscript{\rm 2}$^*$,
    Wenhao Tang\textsuperscript{\rm 1, \rm 3},
    Chao Yu\textsuperscript{\rm 1}$^\dagger$,
    Yu Wang\textsuperscript{\rm 1}$^\dagger$
}
\affiliations {
    \textsuperscript{\rm 1}Department of Electronic Engineering, Tsinghua University\\
    \textsuperscript{\rm 2} Institute for Interdisciplinary Information Sciences, Tsinghua University\\
    \textsuperscript{\rm 3} Tsinghua Shenzhen International Graduate School, Tsinghua University\\
    $^*$ Equal contribution, qiul21@mails.tsinghua.edu.cn \\
    $^\dagger$Corresponding authors: zoeyuchao@gmail.com, yu-wang@tsinghua.edu.cn
}

\begin{document}
\thispagestyle{firstpagehf}
\maketitle

\begin{abstract}
\input{sections/00abs}
\end{abstract}


\section{INTRODUCTION}
\input{sections/10intro}

\section{RELATED WORK}
\input{sections/20related}

\section{PRELIMINARY}
\input{sections/30pre}

\section{METHODOLOGY}

\input{sections/40method_workshop}

\section{EXPERIMENT}
\input{sections/50exp_workshop}

\section{CONCLUSION}
\input{sections/60conclusion}


\bibliography{reference}


\clearpage
\onecolumn
\appendix
\section*{Appendices}
\input{sections/70appendix}

\end{document}

%% file: sections/00abs.tex
Assessing the safety of autonomous driving policy is of great importance, and reinforcement learning (RL) has emerged as a powerful method for discovering critical vulnerabilities in driving policies. However, existing RL-based approaches often struggle to identify vulnerabilities that are both effective—meaning the autonomous vehicle is genuinely responsible for the accidents—and diverse—meaning they span various failure types. To address these challenges, we propose \underline{AED}, a framework that uses large language models (LLMs) to \underline{A}utomatically discover \underline{E}ffective and \underline{D}iverse vulnerabilities in autonomous driving policies. We first utilize an LLM to automatically design reward functions for RL training. Then we let the LLM consider a diverse set of accident types and train adversarial policies for different accident types in parallel. Finally, we use preference-based learning to filter ineffective accidents and enhance the effectiveness of each vulnerability. Experiments across multiple simulated traffic scenarios and tested policies show that AED uncovers a broader range of vulnerabilities and achieves higher attack success rates compared with expert-designed rewards, thereby reducing the need for manual reward engineering and improving the diversity and effectiveness of vulnerability discovery. The implementation can be found on: \url{https://github.com/thu-nics/AED}.

%% file: sections/10intro.tex
Autonomous driving is a safety-critical domain where even minor decision-making flaws can lead to severe accidents~\cite{guo2019safe}. 
One approach to assessing the safety of autonomous driving policies is vulnerability discovery, which systematically uncovers scenarios where autonomous driving policy fails or behaves undesirably~\cite{rastogi2020threats,wan2022too}.
By controlling the behaviors of other traffic participants, vulnerability discovery create adversarial driving scenarios where the tested driving policy is exposed to hazardous conditions. 
Due to the rarity of such failure cases in real-world road tests, automated vulnerability discovery methods have been developed in simulation environments, where reinforcement learning (RL) emerges as an efficient approach that trains adversarial agents to induce accidents of the tested policy~\cite{kuutti2020training, feng2021intelligent, chen2021adversarial, mu2024multi}.


\begin{figure}[t]
  \centering
  \includegraphics[width=\linewidth]{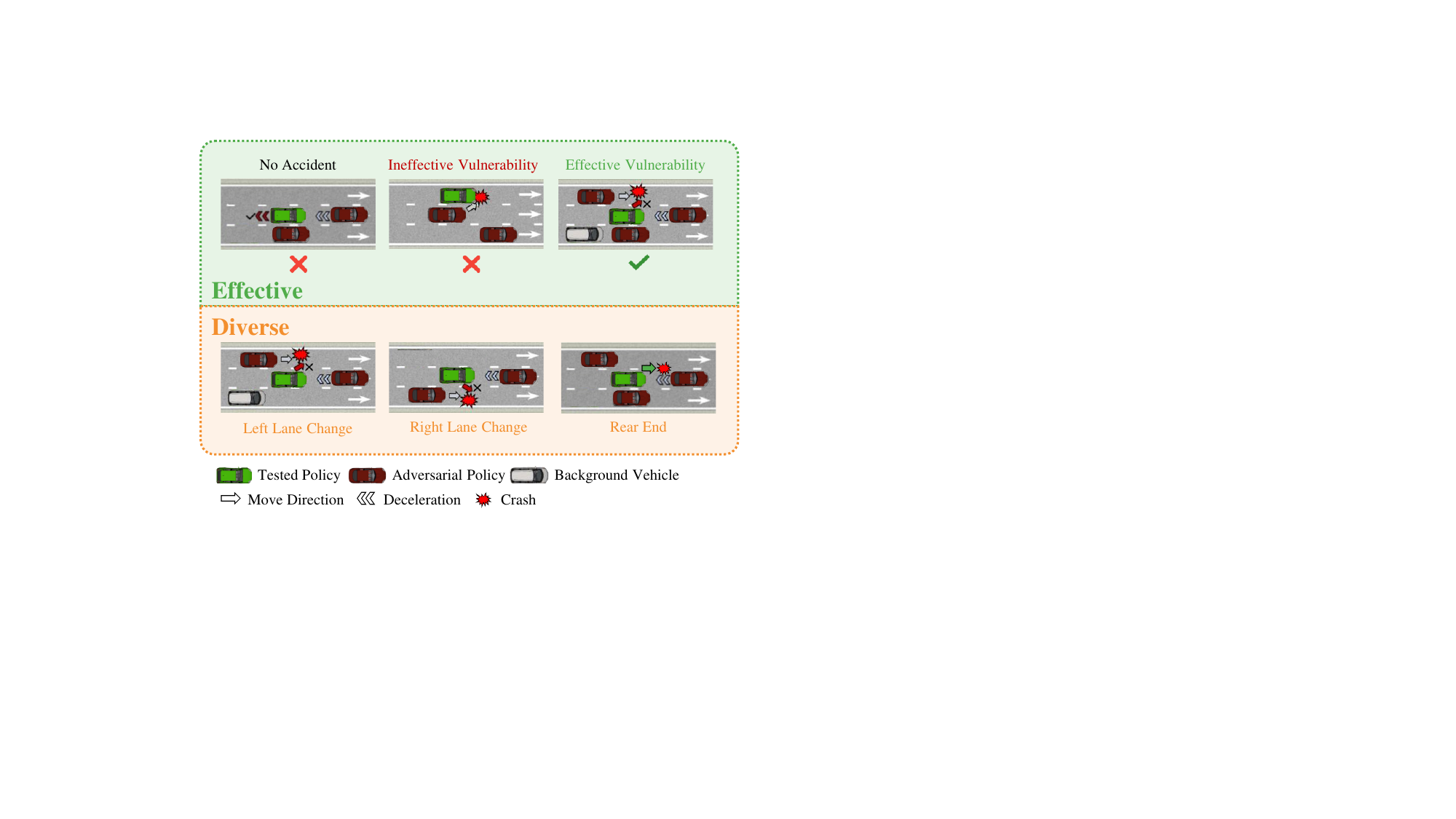}
  \caption{Illustration of effective and diverse vulnerabilities. An effective vulnerability is an accident that is caused by the false decision-making of the tested policy. Diverse vulnerabilities refer to different accident types that are consistent with human understanding and regulations.}
  \label{fig:objective}
\end{figure}

However, there are two key challenges in leveraging RL for vulnerability discovery. 
First, it is hard to automatically identify \textbf{effective} vulnerabilities that require determining accident responsibility. 
When an accident occurs, it is essential to attribute the failure to the autonomous vehicle’s (AV) decision-making or those of other traffic participants. 
An effective vulnerability is one where the tested policy is responsible for the accident, revealing a genuine deficiency in its driving policy \cite{schulman2015trust, corso2019adaptive, mu2024multi}. 
Accident responsibility, however, is complex to assess, often requiring human experts with domain knowledge in real-world settings \cite{shalev2017formal, sun2024acav}.
Similarly, in simulation-based RL training, designing a suitable reward function for responsibility attribution also demands substantial domain expertise \cite{kwon2023reward, ma2023eureka, yu2025largelanguagemodelsincontext}, posing a challenge to the automated discovery of effective vulnerabilities.


The second challenge in RL-based vulnerability discovery is ensuring \textbf{diverse} failure cases. Since a safety-critical flaw can arise from various accident types, it is crucial to comprehensively explore the space of possible failures \cite{corso2019adaptive, koren2020adaptive, mu2024multi}. 
Traditional RL methods, however, tend to converge on the most probable failure case, leading to a lack of diversity in discovered vulnerabilities \cite{eysenbach2018diversity}. Existing approaches to diversity-aware RL generally fall into two types.
The first type is optimizing a general diversity metric alongside task objectives.
However, these methods often lead to vulnerabilities with uninterpretable differences that are not consistent with human understanding and regulations.
The second type is incorporating heuristic-driven diversity measures tailored to specific tasks, which rely heavily on expert knowledge and struggle to generalize across different scenarios. 
These limitations hinder the interpretability and automation of diverse vulnerability discovery, leaving potential safety risks unsolved.



To address the aforementioned challenges, we propose \textbf{\underline{AED}}, a framework that leverages large language models (LLMs) to \underline{\textbf{A}}utomatically discover \underline{\textbf{E}}ffective and \underline{\textbf{D}}iverse vulnerabilities in autonomous driving policies. 
Our framework consists of three key steps. First, an LLM is prompted with the environment and task descriptions to automatically design reward functions for vulnerability discovery. 
Then, to enhance the diversity of vulnerabilities, we use the LLM to consider a broad range of accident scenarios and train adversarial policies for different types in parallel.
Finally, to further improve the effectiveness of the discovered vulnerabilities, we use preference-based reinforcement learning to train a reward model that filters ineffective accidents.
By combining automatic reward design, diverse accident generation, and preference-based effectiveness enhancement, our framework automatically identifies a broad spectrum of effective and diverse vulnerabilities, improving the comprehensiveness of AV safety evaluation.

To evaluate the performance of our framework, we apply our method to discover vulnerabilities of both planning-based policies and learning-based policies across different traffic scenarios, involving different numbers of adversarial vehicle agents. Experiment results show that our method can successfully identify and distinguish a diverse set of different vulnerabilities. Moreover, our method is able to achieve higher attack success rates with automatically generated rewards than policies with expert-designed rewards. These results show that our proposed framework can discover a diverse set of effective vulnerabilities in various tested policies, significantly reducing the reliance on manually designed reward functions and diversity metrics.

In summary, the contributions of this paper are: (1) We propose to use LLMs to automatically discover a diverse set of potential vulnerabilities, enhancing the overall safety of a driving policy. (2) We propose a preference-based learning method to filter ineffective accidents and encourage the identification of effective vulnerabilities. (3) We demonstrate our approach through extensive experiments and analysis across various traffic environments and tested driving policies, demonstrating its robustness and generalizability.

%% file: sections/20related.tex
\begin{figure*}[htp]
  \centering
  \includegraphics[width=\textwidth]{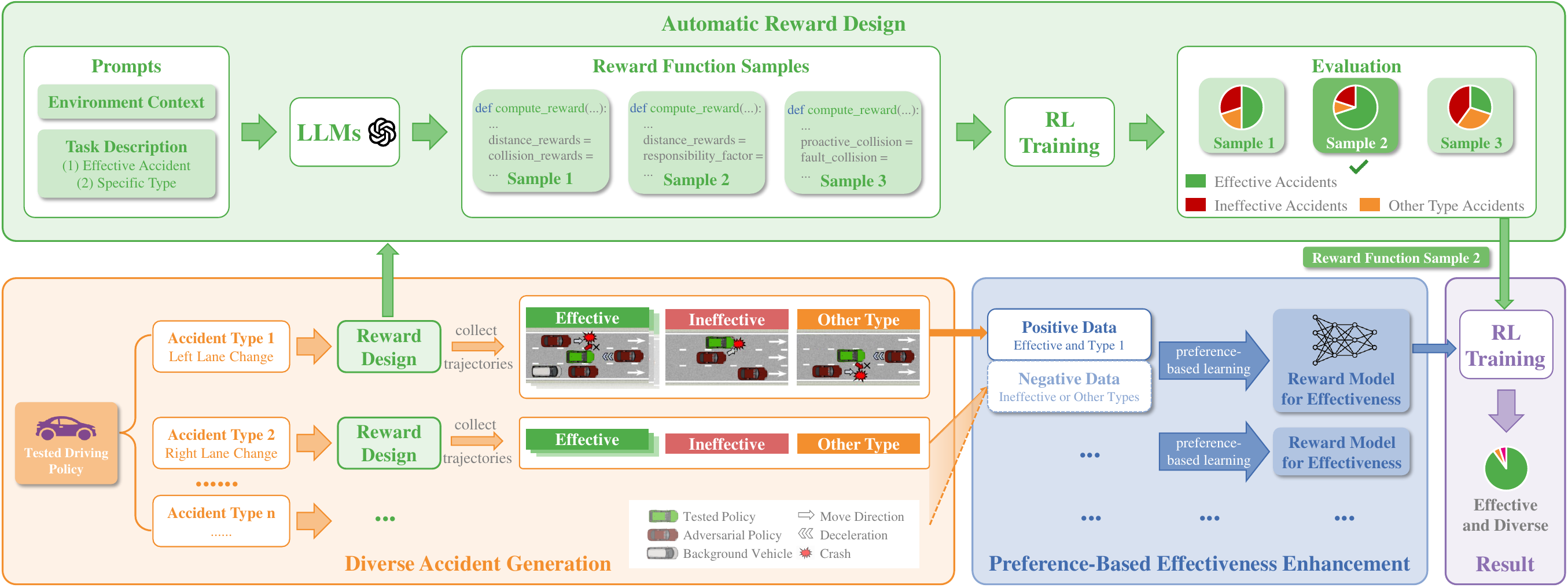}
  \caption{Overview of our proposed AED framework that uses large language models (LLMs) to automatically discover effective and diverse vulnerabilities in autonomous driving policies. Our framework first utilizes an LLM to automatically design reward for vulnerability discovery. Then we use the LLM to consider different accident types and generate a diverse set of accidents in parallel. Finally, we use preference-based learning to enhance the effectiveness of each accident type. Combining the three components leads to our framework for automatic, effective, and diverse vulnerability discovery.}
  \label{fig:overview}
\end{figure*}

\subsection{Diverse Vulnerability Discovery}

Adversarial RL has been widely used to generate safety-critical scenarios that reveal vulnerabilities in autonomous driving policies.
Prior works use Deep Q-Network \cite{mnih2015human} to generate corner cases \cite{karunakaran2020efficient, feng2021intelligent, sun2021corner}, and apply DDPG \cite{lillicrap2015continuous} to create challenging lane-change conditions\cite{chen2021adversarial}. Adaptive Stress Testing \cite{corso2019adaptive} incorporates a Responsibility-Sensitive Safety (RSS) \cite{shalev2017formal} reward into TRPO \cite{schulman2015trust} to classify AV failures. Multi-agent extensions such as MADDPG \cite{lowe2017multi} and MAPPO \cite{yu2022surprising} have been applied to coordinate multiple adversarial vehicles \cite{wachi2019failure, mu2024multi}.
However, these methods rely on crafted reward functions, which require domain expertise to balance responsibility attribution and behavioral realism. 

To improve the diversity of discovered vulnerabilities, existing methods can be categorized into general and heuristic methods. General methods optimize information-theoretic objectives \cite{eysenbach2018diversity} or apply trajectory similarity metrics \cite{liu2012similarity, corso2019adaptive}, but often lack interpretability in real-world scenarios. Heuristic methods incorporate expert knowledge, such as clustering \cite{chen2021adversarial} or intrinsic novelty reward \cite{mu2024multi} to promote rare failures, but remain dependent on handcraft designs and domain-specific assumptions. Our work addresses these limitations by using LLMs to automatically generate diverse reward functions tailored to distinct vulnerability types, improving adaptability while reducing the reliance on manual heuristics.

\subsection{Reward Design with LLMs}

Reward design is a fundamental challenge in reinforcement learning, often requiring extensive domain expertise and iterative trial-and-error~\cite{sutton2018reinforcement,singh2009rewards}. 
Recent advances leverage LLMs to automate reward design by exploiting their capabilities in reasoning, planning, and code generation~\cite{shinn2023reflexionlanguageagentsverbal,du2023guidingpretrainingreinforcementlearning,carta2024groundinglargelanguagemodels}.
Some studies use LLMs to interpret environment and agent behaviors and assign intrinsic rewards based on scenario understanding~\cite{kwon2023reward,wu2024readreaprewardslearning,chu2023accelerating,wang2024sociallymorallyawarerl}, while others focus on generating executable reward code~\cite{ma2023eureka,yu2025largelanguagemodelsincontext,yu2023languagerewardsroboticskill,li2024automcrewardautomateddense}. Notably, Eureka~\cite{ma2023eureka} proposes an automatic self-reflective framework where LLMs generate and iteratively refine reward functions using environment code and sparse human-designed evaluation rewards. ICPL~\cite{yu2025largelanguagemodelsincontext} extends this to tasks without predefined rewards, enabling preference-based learning from human feedback. Our work extends ICPL to multi-agent traffic scenarios, where reward design must consider agent interactions and responsibility attribution.


\subsection{Preference-Based Reinforcement Learning}

Preference-based RL learns directly from expert preference rather than handcrafted numeric rewards, reducing the reliance on domain-specific knowledge. Prior works \cite{akrour2012april, wilson2012bayesian, wirth2016model, mathewson2017actor, lee2021pebble, myers2022learning, christiano2017deep} explore RL from human feedback via trajectory preference queries or score ranking. However, querying human labels during online training is often costly and slow.
To address this limitation, recent studies have proposed RL from AI feedback \cite{bai2022constitutional}. Some leverage LLMs to provide preference based on vision-based trajectory or text descriptions of states and actions \cite{wang2024rl, wang2025prefclm, klissarov2023motif, chu2023accelerating, tu2024online}. Our work automatically generates preference data pairs from simulation outcomes produced by parallel adversarial policies, enabling scalable preference learning without numerous labeling while capturing diverse vulnerabilities in multi-agent interactions.

%% file: sections/30pre.tex

\subsection{Partially Observed Markov Decision Process}

Our MARL framework is modeled as a Partially Observed Markov Decision Process (POMDP). defined by a tuple $\mathcal{M} = (\mathcal{N}, \mathcal{S}, \mathcal{A}, \mathcal{O}, \mathcal{P}, \mathcal{R}, \gamma)$, where $\mathcal{N}=\{1, 2, \cdots, N\}$ is the set of agents, $\mathcal{S}$ is the state space covering all possible states of the N agents, $\mathcal{A}$ is the shared action space, $\mathcal{O}$ denotes each agent's observation space, $\mathcal{P}$ is the transition probability function, $\mathcal{R}$ is the joint reward function with $\gamma$ is the reward discount factor. At each step, agent $i$ receives an observation $o_i = \mathcal{O}(s_i)$ where $s_i \in \mathcal{S}$ is the local state. The agent then takes an action $a_i$ according to the shared policy $\pi_\theta(a_i | o_i)$, where $\theta$ represents the shared parameters. The next state progresses to $s_i'$ according to the transition probability $p(s_i' | s_i, a_i) \in \mathcal{P}$ and the agent $i$ receives a reward $r_i = \mathcal{R}(s_i, a_i)$. For homogeneous agents, each agent $i$ optimizes the discounted cumulative rewards $J(\pi) = \mathbb{E}_{\tau \sim \pi} \left[ \sum_t \gamma^t r_t \right]$, where a trajectory $\tau$ denotes a state-action-reward triplets sequence, i.e., $\tau = \{(s_t, a_t, r_t)\}_{t}$.

\subsection{Diverse and Effective Vulnerability Discovery}

Our objective is to discover diverse and effective vulnerabilities in a fixed, pre-trained autonomous driving policy by training adversarial policies within the previously defined POMDP framework. While the tested policy remains fixed, adversarial vehicles are optimized to induce failure scenarios that expose weaknesses in the tested policy. 
Specifically, the adversarial policy is optimized to maximize the rate of effective vulnerability, defined as those for which responsibility can be attributed to the tested policy. Diversity is assessed by the number of distinct vulnerability types, and effectiveness is assessed by the proportion of vulnerabilities discovered arising from wrong decisions of the tested driving policy. Formal definitions and evaluation procedures for these two metrics are provided in Sec.~\ref{EXP: Diversity} and Sec.~\ref{EXP: Effectiveness}.

%% file: sections/40method_workshop.tex


To automatically find effective and diverse vulnerabilities, we propose the AED framework that consists of three components including automatic reward design, diverse accident generation, and preference-based effectiveness enhancement. The overview of our framework is shown in Fig.~\ref{fig:overview}.
The first automatic reward design component prompts an LLM with the environment and task descriptions to automatically generate reward function samples and select the best one. 
Then, to enhance the diversity of vulnerabilities, we propose a diverse accident generation component that prompts the LLM to consider a broad range of accident types and trains adversarial policies for different types in parallel.
Finally, to enhance the effectiveness of each vulnerability, we use preference-based learning to train a reward model that filters ineffective accidents and other accident types.
Combining the three components lead to our framework that automatically discovers effective and diverse vulnerabilities.

\subsection{Automatic Reward Design}
\label{method: automatic reward design}


To enable automatic reward design for vulnerability discovery, we extend the In-Context Preference Learning (ICPL) framework \cite{yu2025largelanguagemodelsincontext} to leverage the code-synthesis and in-context learning capabilities of LLMs. In our adaptation, the LLM generates task-specific reward functions based on environment descriptions and human preferences, guiding the discovery of effective accidents in a self-improving manner.

The LLM first receives an environment description and a detailed task specification, then generates $K$ executable reward functions. The environment text outlines the scenario and relevant code-level details, such as the agent’s observation/action space and key parameters or functions useful for reward construction. The specific details of the task description are provided in Sec.~\ref{method:Diverse_Generation}. Each reward function is used to train an agent, resulting in $K$ agents, which are evaluated using human-written criteria for identifying effective accidents. To ensure comprehensive coverage, we also generate accident videos for human evaluation. Based on both metrics and visual inspection, evaluators select the best reward functions. The selected best function typically yields high accident effectiveness across scenarios and serves as a basis for further refinement.


\subsection{Diverse Accident Generation}
\label{method:Diverse_Generation}
To enhance diversity in vulnerability discovery, the adversarial policy should capture distinct characteristics of various weaknesses in the tested policy. We introduce a Diverse Accident Generation component, which prompts LLMs to generate a set of reward functions that promote vulnerability diversity. By considering a wide range of accident types, LLMs automatically design multiple reward functions in parallel, each tailored to induce specific failure cases.


The process begins with the LLM generating a task description for a given accident type based on environment specifications. This description includes a high-level summary of the targeted accident scenario, concrete examples, justifications for attributing responsibility to the tested agent, and criteria for categorizing scenarios within this accident type. Using the component outlined in Sec.~\ref{method: automatic reward design}, the LLM then autonomously generates reward functions corresponding to different accident types in parallel.

Once reward functions are generated, they are integrated into the training of adversarial policies, each optimized for a specific vulnerability of the tested policy. These adversarial policies are then used to collect trajectory data, creating a structured dataset covering diverse failure cases. This dataset serves as a foundation for further analysis and improvement of the tested policy, ensuring a comprehensive evaluation of its robustness across various accident scenarios. 

The prompts used to guide the LLM in generating diverse accident types, along with the corresponding environment and scenario descriptions, are provided in Appendices.

\begin{figure}[t]
    \centering
    \includegraphics[width=\linewidth]{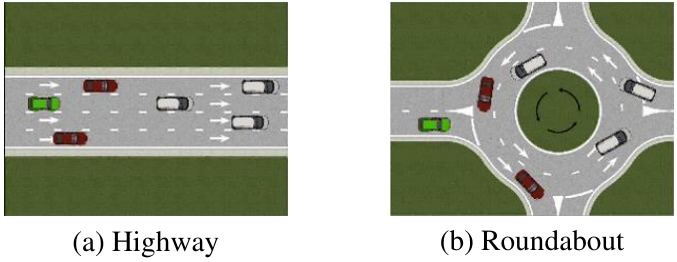}
    \caption{Demonstration of ``Highway'' and ``Roundabout''.}
    \label{fig:env_demonstration_exp}
\end{figure}

\begin{figure*}[ht]
  \centering
  \includegraphics[width=0.9\textwidth]{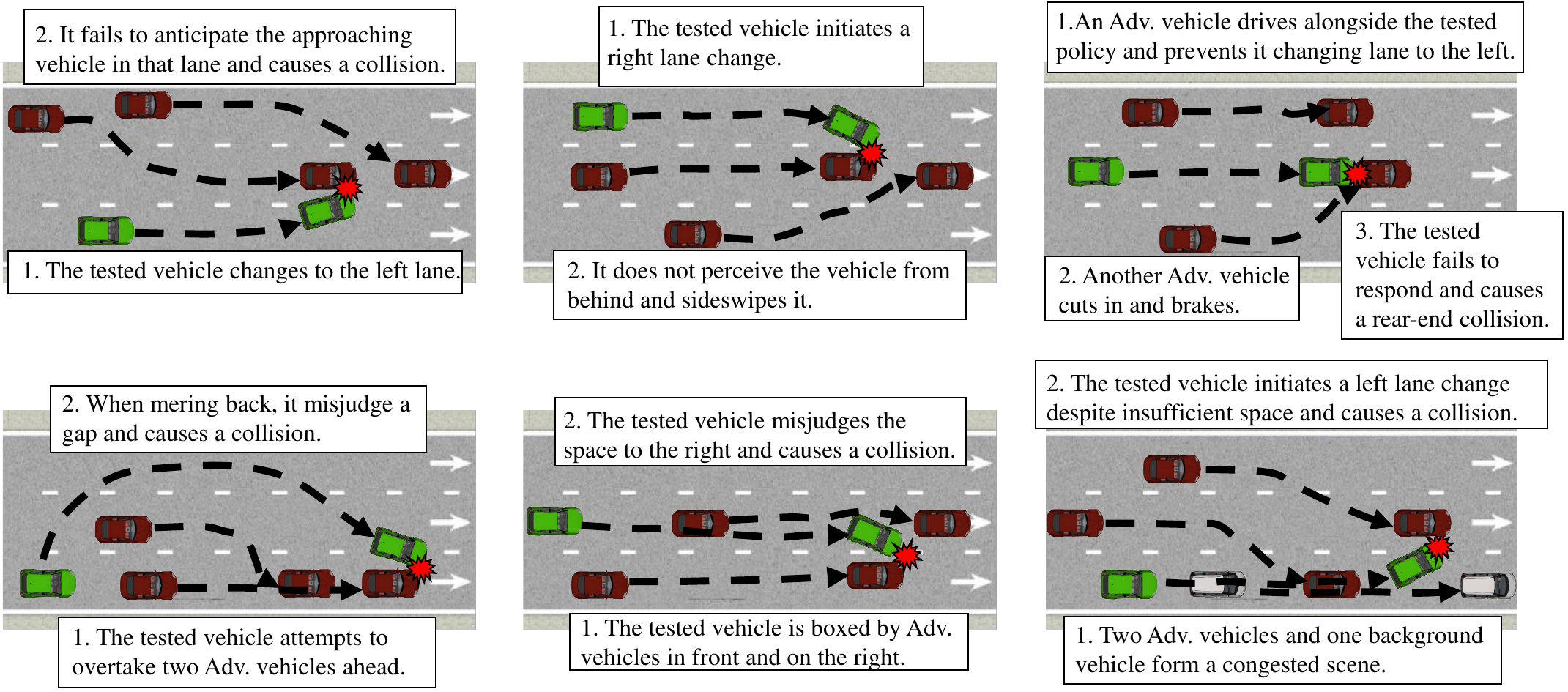}
  \caption{Examples of distinct vulnerability types discovered by AED. Adv. is abbreviation for             ``adversarial''.} 
  \label{fig:visualize_traj_exp}
\end{figure*}

\begin{figure*}[ht]
    \centering
    \subfloat{\includegraphics[width=0.25\textwidth]{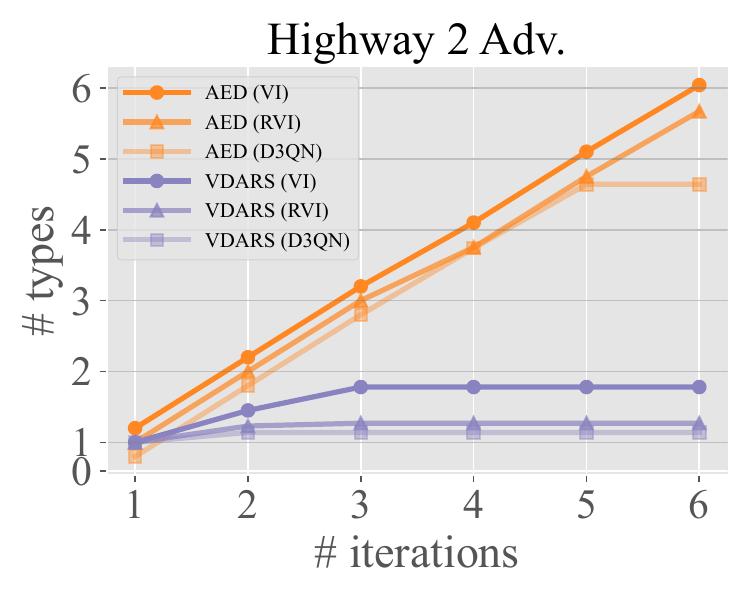}}
    \subfloat{\includegraphics[width=0.25\textwidth]{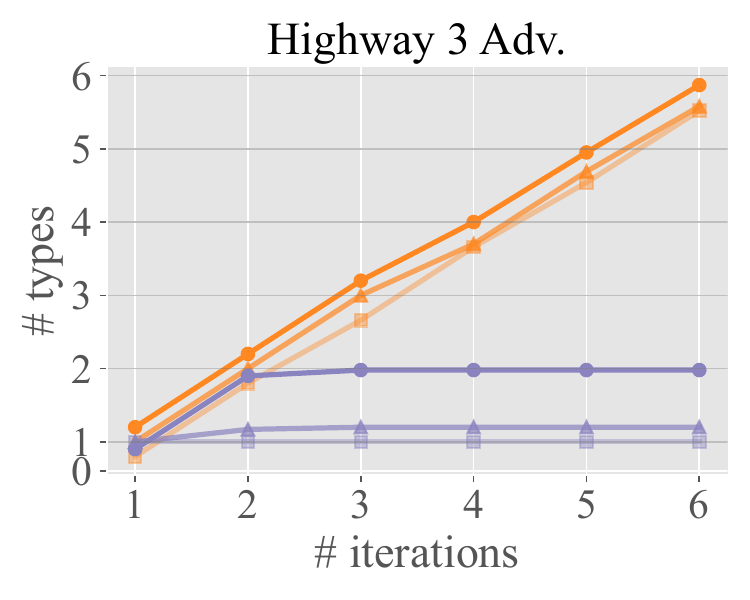}} 
    \subfloat{\includegraphics[width=0.25\textwidth]{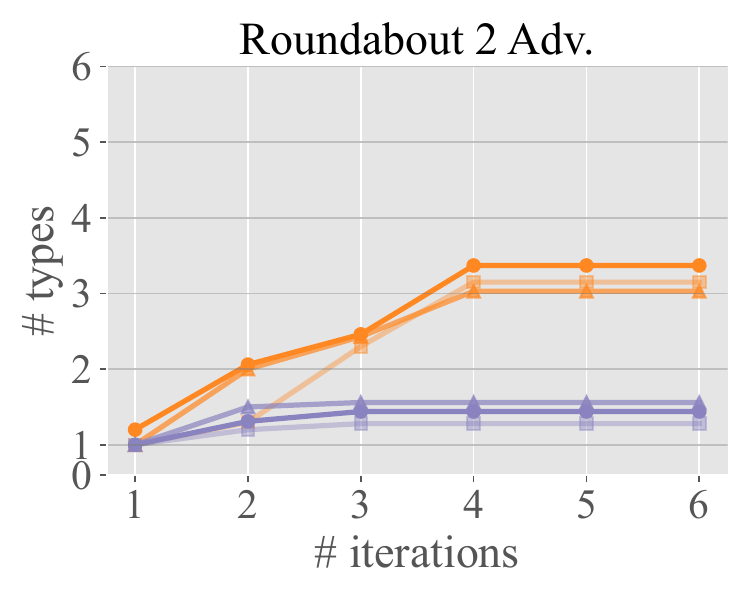}}
    \subfloat{\includegraphics[width=0.25\textwidth]{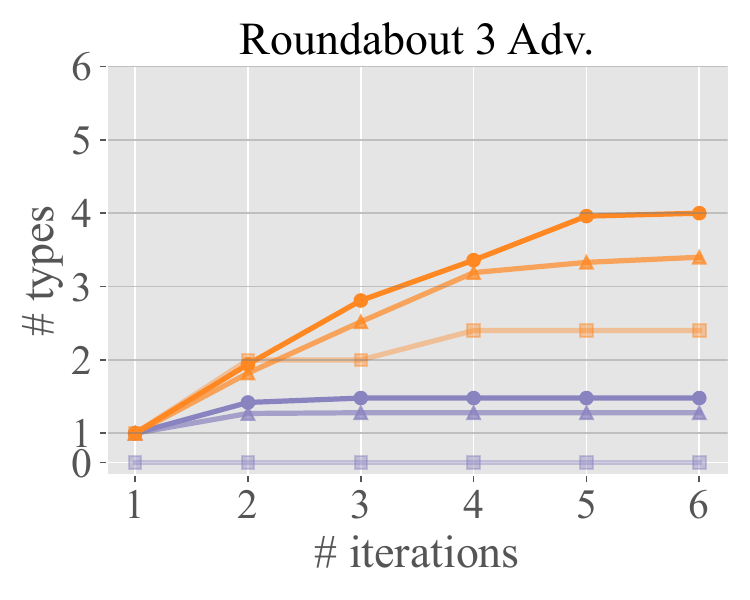}}
    \caption{Number of vulnerability types discovered across different environments, tested policies, and numbers of adversarial vehicle agents. Each line represents the cumulative number over six training iterations, where the first discovered type is counted as 1, and subsequent new types are weighted by their relevant frequency among effective failures. AED consistently discovers more diverse vulnerabilities than VDARS.}
    \label{fig:diverse}
\end{figure*}

\subsection{Preference-Based Effectiveness Enhancement}


Reward functions derived from Sec.~\ref{method:Diverse_Generation} for a specific accident type are often imperfect, sometimes capturing unintended accident types or ineffective scenarios. For instance, the reward function for Left Lane Change (Accident Type 1) may capture Right Lane Change (Accident Type 2) and Adversarial Vehicle cut in mistakenly (Accident Type 3). To improve the effectiveness and distinguish diverse accident types, we propose a preference-based effectiveness enhancement component to refine the LLMs-driven reward function. The key idea is to learn a new reward function $\hat{r}$ that prefers trajectories corresponding to the intended accident types over other types. Specifically, $\hat{r}$ is trained to assign higher cumulative rewards to Type 1 than to those from Type~2 or Type 3.

To construct training data for preference learning, we leverage trajectory data generated in parallel for different accident types using their respective LLMs-driven reward functions. We consider the Effective and Type 1 dataset as Positive Data $\sigma^1$, and use a combination of the Ineffective or Other Types (Type 2 and Type 3) as Negative Data $\sigma^2$. 

The probability of preferring a trajectory $\sigma^i$ as the intended accident type is assumed to depend exponentially on the accumulated latent reward over the trajectory's state-action sequence $(o_t^{i}, a_t^{i})$ :

\begin{equation}
    \hat{P}[\sigma^1 \succ \sigma^2] = \frac{\exp \sum_t \hat{r}(o_t^1, a_t^1)}{\exp \sum_t \hat{r}(o_t^1, a_t^1) + \exp \sum_t \hat{r}(o_t^2, a_t^2)}. 
\end{equation}

To optimize the reward function $\hat{r}$, we minimize the cross-entropy loss between these predictions and the intended preference. The Positive Data and Negative Data are stored in a database $\mathcal{D}$ of triples $(\sigma^1, \sigma^2, \mu)$, where $\mu$ is a judgment label (0 or 1) to indicate whether $\sigma^1$ is the preferred accident type. The loss function is defined as:

\begin{align}
    \text{loss}(\hat{r}) = - \sum_{(\sigma^1, \sigma^2, \mu) \in \mathcal{D}} & \mu \log \hat{P}[\sigma^1 \succ \sigma^2] + \notag \\
    & (1-\mu) \log (1-\hat{P}[\sigma^1 \succ \sigma^2]).
\end{align}
    
Since $\mu$ assigns full probability mass to the Positive Data, which means $\mu=1$ across the dataset, the loss simplifies to:

\begin{equation}
    \text{loss}(\hat{r}) = - \sum_{(\sigma^1, \sigma^2, \mu) \in \mathcal{D}} \log \hat{P}[\sigma^1 \succ \sigma^2].
\end{equation}

In this way, the reward function $\hat{r}$ learns to favor Type 1 accident trajectories while effectively distinguishing them from other types. We then incorporate $\hat{r}$ and the original LLMs-driven reward function into RL training. If $\hat{r}$ identifies trajectories as Type 2, the training will be penalized. This guides RL training to prioritize Accident Type 1 and ultimately improve the attack success rate for this category.




%% file: sections/50exp_workshop.tex


We evaluate AED in two driving environments ``Highway'' and ``Roundabout'' using the Highway simulator \cite{leurent2018environment}, as shown in Fig.~\ref{fig:env_demonstration_exp}. We consider three types of tested driving policy: two Planning-based methods, Value Iteration (VI) \cite{bellman1966dynamic} and Robust Value Iteration (RVI) \cite{nilim2004robust}, and one RL-based method, Dueling Double Q-Network (D3QN) \cite{wang2016dueling}. To account for the complexity of traffic interactions, we conduct experiments with two and three adversarial vehicles in each environment.

\subsection{Implementation Details}
\label{EXP: Implementation Details}
We categorize vulnerabilities into three types based on the relative positions of the tested and adversarial vehicles: \textbf{Left Lane Change (Left)}, \textbf{Right Lane Change (Right)}, and \textbf{Rear End (Rear)} collision scenarios. These categories correspond to different failure-inducing situations. For later diversity analysis, we further divide these categories into finer-grained subtypes based on common real-world traffic accident patterns. Each adversarial policy is trained using the MAPPO algorithm \cite{yu2022surprising} for 10 million steps. We utilize GPT-4o and DeepSeek-R1 for diverse accident generation. Detailed environment settings and training parameters are provided in the Appendices.

We compare the performance of our method with the following baselines:
\begin{itemize}
    \item \textbf{VDARS} \cite{mu2024multi}: A multi-agent RL framework that employs handcrafted reward functions to learn the adversarial policy for vulnerability discovery.
    \item \textbf{Parallel Eureka}: An ablated version of our method without the preference-based effectiveness enhancement module. This can also be regarded as a parallel version of Eureka for different vulnerability types.
\end{itemize}


\subsection{Evaluation of Diversity}
\label{EXP: Diversity}

To evaluate AED's ability to discover diverse vulnerability scenarios, we measure the number of distinct vulnerability types identified across multiple training iterations. We divide the three high-level collision categories into six fine-grained subtypes to enable more precise diversity assessment. Illustrative examples discovered in the Highway environment are shown in Fig.~\ref{fig:visualize_traj_exp}. A complete set of definitions and additional illustrative examples can be found in the Appendices. For each setting---defined by the environment (Highway and Roundabout), the tested policy (VI, RVI, and D3QN), and the number of adversarial vehicle agents (2 and 3)---we conduct six independent training iterations with different random seeds. In each iteration, we collect all effective failure cases. In the first iteration, we identify the most frequently occurring effective vulnerability type and assign it a count of 1. For each subsequent iteration, we exclude all previously discovered types and select the most frequent type among the remaining ones. Instead of simply incrementing the count by one, we add its proportion among all effective failures in that iteration to the cumulative count. This strategy ensures that each iteration contributes at most one new type, and the final diversity score reflects both the variety and prominence of distinct vulnerabilities uncovered. In VDARS, all iterations use the same handcrafted reward function. In contrast, AED trains the adversarial policy in each iteration with a different LLMs-generated reward function targeted at distinct failure types. 

As shown in Fig.~\ref{fig:diverse}, AED consistently discovers more diverse vulnerability types than VDARS. Across all settings, AED reaches or approaches the maximum of six types within six iterations, while VDARS typically uncovers only two or three. This discrepancy highlights the limited exploration capability of a single handcrafted reward and AED's strength in promoting broader vulnerability discovery through multiple LLMs-generated reward functions.

\subsection{Evaluation of Effectiveness}
\label{EXP: Effectiveness}
We evaluate the effectiveness of the proposed preference-based enhancement module through both qualitative and quantitative analyses. These evaluations demonstrate its utility by: (1) assessing the learned reward model's ability to distinguish goal-aligned vulnerabilities from unrelated failures, and (2) measuring its impact on the overall effective vulnerability rate when integrated into RL training.

\subsubsection*{Qualitative Evaluation} 
\begin{figure}[t]
  \centering
  \includegraphics[width=\linewidth]{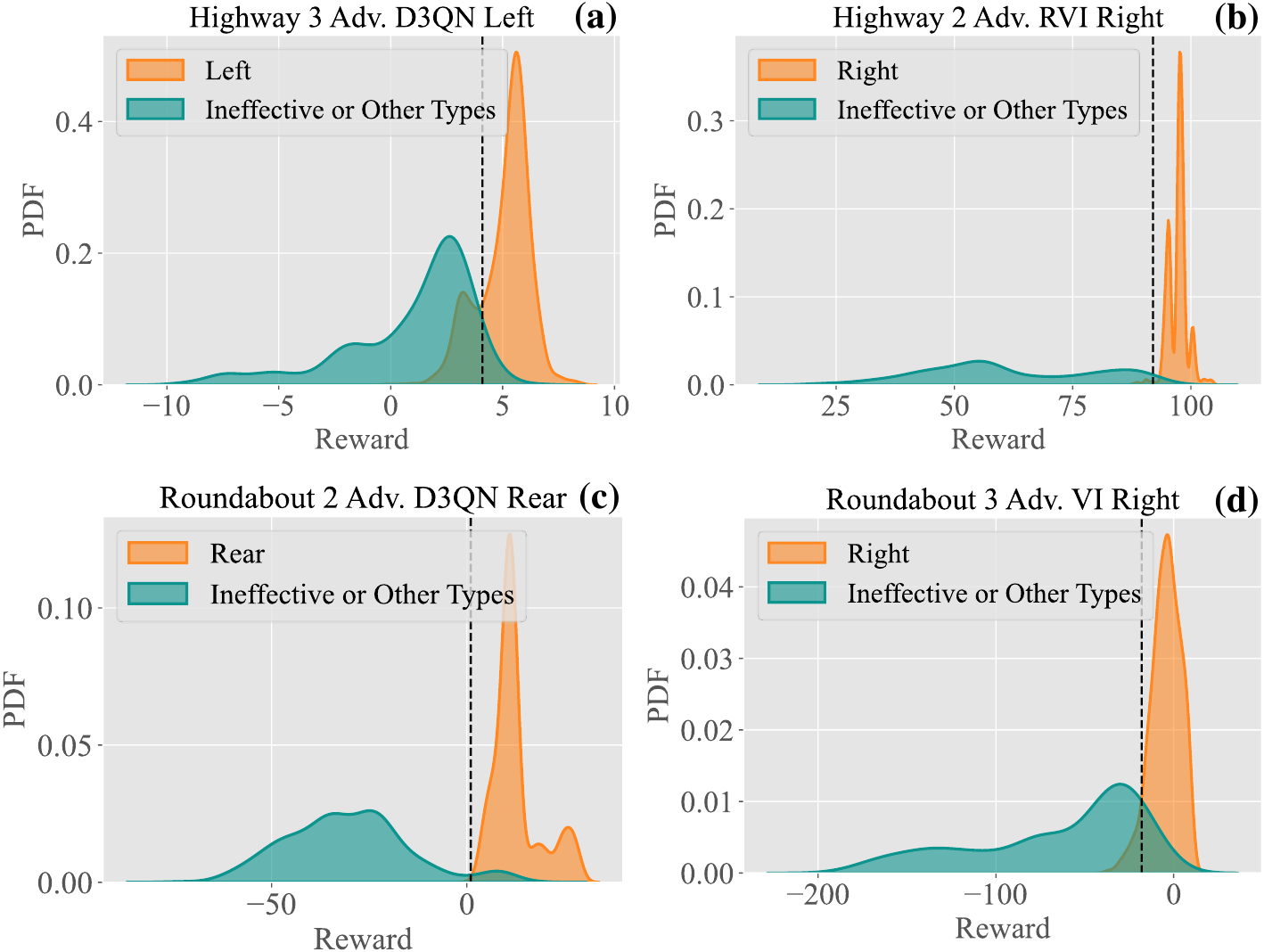}
  \caption{Probability density distribution of reward for accident trajectories across different environments and policy configurations. In all cases, the learned reward model assigns higher scores to goal-aligned failures (orange), effectively separating them from unrelated accidents (green).}
  \label{fig:rew_dist}
\end{figure}

We first illustrate the reward model's discriminative capability using a representative case from the Highway environment with three adversarial vehicle agents and a D3QN tested policy. When trained with a LLMs-generated reward function targeting Left Lane Change (Left) collisions, the resulting accidents include both targeted Left collisions, unintended Rear-End collisions, and other ineffective failures. After applying preference-based reward learning, the learned model is evaluated on different accident trajectories. As shown in Fig.~\ref{fig:rew_dist}(a), it assigns higher reward scores to Left collisions (orange) than to unrelated accidents (green), demonstrating clear separation. Similar reward distributions are observed across other environments and policies (Fig.~\ref{fig:rew_dist}(b)(c)(d)), further validating the model's ability to distinguish goal-aligned vulnerabilities from irrelevant ones.

\begin{figure*}[ht]
  \centering
  \includegraphics[width=0.9\textwidth]{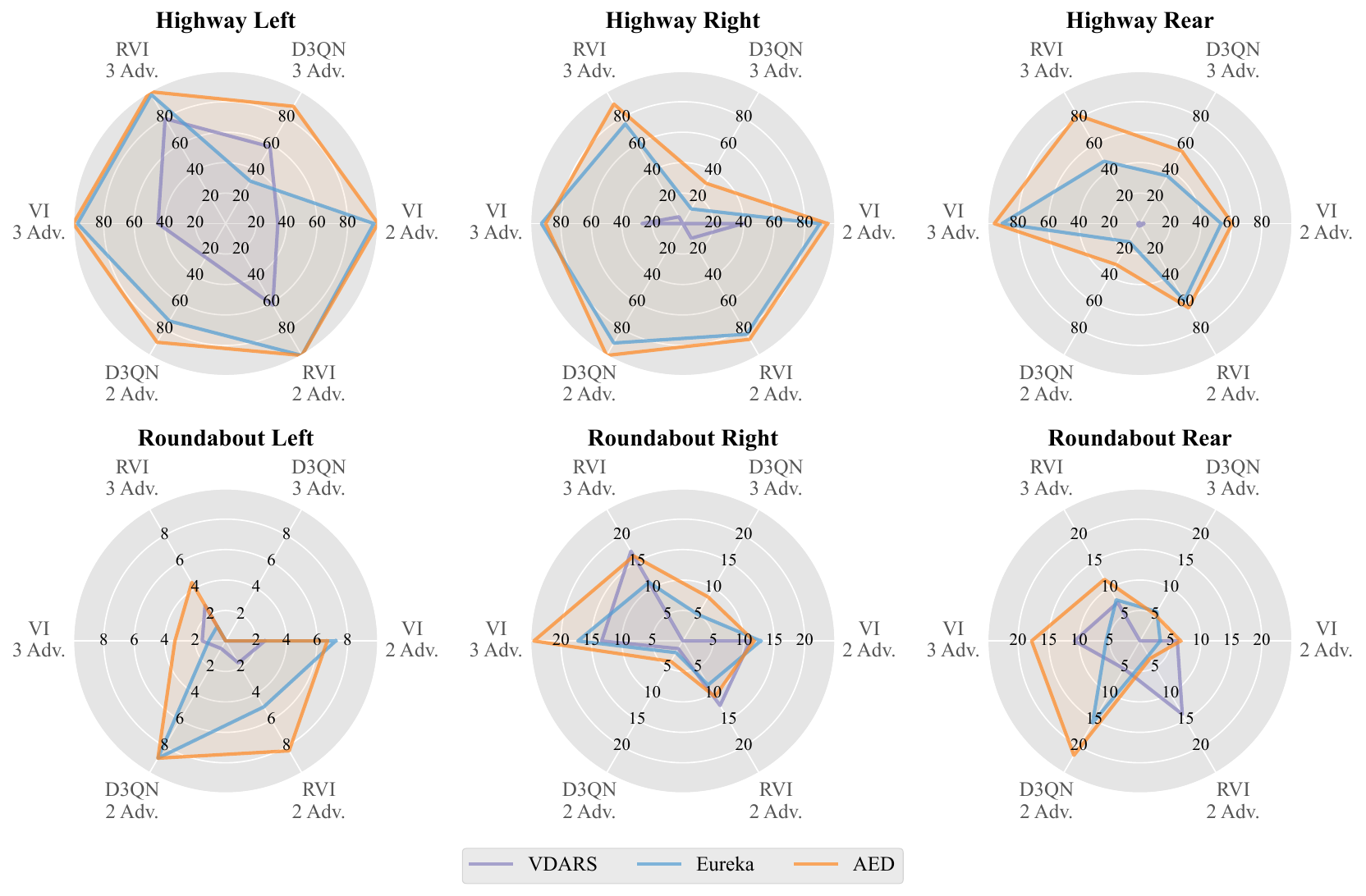}
  \caption{Effective Vulnerability Rates (\%) across different environments, tested policies, and numbers of adversarial vehicle agents. Each radar chart corresponds to one environment with one high-level accident category. Adv. is abbreviation for ``adversarial''. AED consistently achieves the highest effective vulnerability rates.} 
  \label{fig:radar_effectivess}
\end{figure*}

\begin{figure*}[t]
  \centering
  \includegraphics[width=0.95\textwidth]{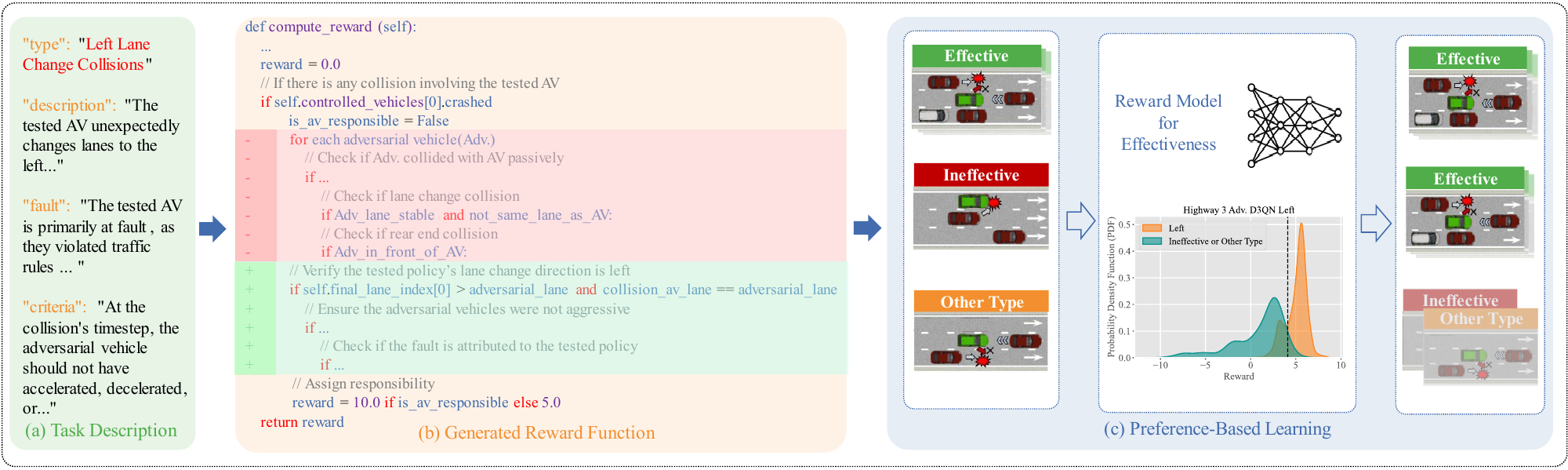}
  \caption{Case study demonstrating the AED framework. (a) Task description provided to LLMs specifies the targeted Left Lane Change vulnerability, scenario details, and responsibility criteria. (b) The LLMs-generated reward function verifies the tested policy's maneuver, ensures non-aggressive behavior from adversarial vehicles, and assigns responsibility accordingly. Differences from the expert-crafted reward are highlighted using a diff-style format. (c) The preference-based reward model differentiates targeted failure scenarios from unintended ones.} 
  \label{fig:example}
\end{figure*}

\subsubsection*{Quantitative Evaluation} To quantify the benefit of preference-based enhancement, we measure the \textbf{effective vulnerability rate}---defined as the percentage of simulation runs where the adversarial policy successfully triggers a specific failure type attributable to the tested policy. We compare the full AED framework against a baseline ablation, Parallel Eureka, which uses LLMs-generated reward functions without the final enhancement module. Parallel Eureka can be seen as a variant of VDARS, replacing a fixed handcrafted reward function with static LLMs-generated ones. In AED, if the learned reward model assigns a score below a predefined threshold (black dashed lines in Fig.~\ref{fig:rew_dist}), the corresponding LLMs-generated reward is suppressed to zero, thus filtering out off-target behaviors.

The effective vulnerability rate reported corresponds to the maximum among all fine-grained subtypes in each high-level collision category. Results for all subtypes are provided in the Appendices. As shown in Fig.~\ref{fig:radar_effectivess}, AED consistently outperforms or matches Parallel Eureka in nearly all settings. The improvement is especially notable where the initial LLMs-generated reward is suboptimal. For example, in the Highway environment with a D3QN-driven tested policy and three adversarial vehicles, AED increases the effective Left collision rate from $32.22\%$ to $88.89\%$. This substantial improvement demonstrates the impact of preference-based enhancement in guiding adversarial vehicles to consistently induce failure scenarios that the tested policy struggles with. Even in cases where Parallel Eureka already performs well, AED maintains or improves performance, indicating that the refinement module enhances reward fidelity without degrading strong initial signals.

\subsection{Overall Performance}
AED consistently outperforms VDARS in both diversity and effectiveness. It approaches a maximum of six distinct vulnerability types in most settings. It also consistently achieves significantly higher effective vulnerability rates (Fig.~\ref{fig:radar_effectivess}). 
These gains stem from AED's design: its parallel accident generation enables broader exploration of failure modes, while the preference-based enhancement module refines noisy LLMs-generated rewards into precision training objects. Together, they make AED a more comprehensive framework for adversarial vulnerability discovery.






\subsection{Interpretability and Discussion}
To illustrate how AED transforms high-level vulnerability descriptions into effective adversarial policies, we present a case study in Fig.~\ref{fig:example}, targeting Left Lane Change collisions in the Highway environment with a D3QN-driven tested policy and three adversarial vehicles. 

\subsubsection*{Task Specification and Reward Generation}
We begin by prompting the LLMs with relevant environment context (e.g., road layout, lane width, vehicle length, and action set) and task description, including the targeted vulnerability type (``Left Lane Change Collisions''), expected behaviors (e.g., ``the tested AV unexpectedly changes lanes to the left...''), and criteria for assigning responsibility. Based on this, the LLM generates $K$ ($K=6$ in our experiments) executable reward functions. Each candidate function is used to train an adversarial policy, and the best-performing one---measured by the proportion of Left Lane Change collisions in simulation runs---is selected.

\subsubsection*{Reward Function Analysis}
The selected reward function is shown in Fig.~\ref{fig:example}(b). It consists of modular checks for: (1) whether the tested vehicle performed a left lane change before collision, (2) whether adversarial vehicles behaved non-aggressively (e.g, within steering angle and acceleration constraints), and (3) whether the collision can be attributed to the tested vehicle. The differences between the LLMs-generated reward and the expert-crafted version are visualized using a diff-style format. While both functions share a similar core structure---detecting collisions and evaluating adversarial passiveness---the expert reward relies on a simplified branching logic that only distinguishes between lane change and rear-end collisions. In contrast, the LLMs-generated reward introduces two key advantages. First, it explicitly encodes the targeted collision type (Left Lane Change) as part of the reward condition. Second, it examines both the current and previous actions of the AV to more accurately determine responsibility. These improvements ensure that rewarded collisions are not only aligned with the intended failure type, but also effectively attributed to decision-making flaws in the tested policy.

\subsubsection*{Preference-based Enhancement}
Initial adversarial training with the LLMs-generated reward function yields a suboptimal distribution of accident types: Left Lane Change collisions $(36.67\%)$, Rear-End collisions $(1.00\%)$, and ineffective accidents $(62.33\%)$. As shown in Fig.~\ref{fig:example}(c), the preference-based model corrects this by assigning higher rewards to targeted Left Lane Change collisions and lower rewards to unintended or ineffective accidents. During RL training, the preference model is integrated with the original LLMs-generated reward function. If the preference score for a trajectory during training falls below a predefined threshold, the LLMs-generated reward is suppressed to zero. This filtering mechanism shifts the training focus towards high-quality, goal-aligned failures, resulting in a substantial increase in effective Left Lane Change collisions to $88.89\%$.

%% file: sections/60conclusion.tex
We propose AED, an LLMs-based framework that uses reinforcement learning to automatically discover effective and diverse vulnerabilities of autonomous driving policies. We use an LLM to automatically design reward functions without the need of human experts. To enhance the diversity of vulnerabilities, we use the LLM to consider a broad range of potential accidents and train adversarial policies using RL in parallel. We further improve the effectiveness of each vulnerability by training a reward model with preference-based learning to filter ineffective or other types of accidents. Experiments on various driving scenarios with both planning-based policies and learning-based policies show that AED discovers a diverse range of vulnerabilities and achieves higher attack success rates compared to existing methods, showing the potential of using LLMs to reduce the need for manual reward engineering and improve the comprehensiveness of autonomous driving safety evaluation.

%% file: sections/70appendix.tex







\subsection*{MAPPO Training}
\subsubsection{Observation and Action Space}
In our experiments, all agents share the same observation and action space. Each agent receives an observation vector $o_i = [k_0^i, k_1^i, k_2^i, k_3^i, k_4^i]$, where each $k_j^i$ represents the state of one of the five nearest neighboring vehicles. Each $k$ consists of five features $k = [p, x, y, v_x, v_y]$ as detailed in Table.~\ref{tab:obs_space}. Each agent selects from a discrete action space consisting of five maneuver options, as shown in Table.~\ref{tab:action_space}.



\begin{table}[htp]
    \centering
    \begin{minipage}[t]{0.48\textwidth}
        \centering
        \footnotesize
        \input{tabs/obs_space}
        \caption{Observation space in MAPPO training.}
        \label{tab:obs_space}
    \end{minipage}
    \hfill
    \begin{minipage}[t]{0.48\textwidth}
        \centering
        \footnotesize
        \input{tabs/action_space}
        \caption{Discrete action space available to agents.}
        \label{tab:action_space}
    \end{minipage}
\end{table}

\subsubsection{Environment Setting}
\begin{figure}[H]
    \centering
    \includegraphics[width=0.6\textwidth]{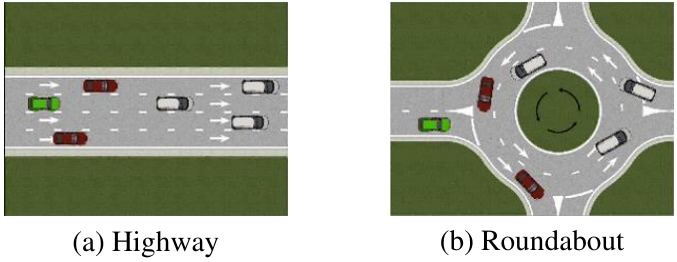}
    \caption{Demonstration of ``Highway'' and ``Roundabout''.}
    \label{fig:env_demonstration}
\end{figure}

To account for diverse and realistic traffic scenarios, we conduct experiments in two representative simulation environments: ``Highway'' and ``Roundabout' as demonstrated in
Fig.~\ref{fig:env_demonstration}. \textit{Highway} scenarios involve high-speed interactions, frequent lane changes, rear-end collisions and overtake collisions, which are typical in structured multi-lane traffic. \textit{Roundabout} scenarios involve frequent turning, right-of-way decisions and merging behaviors. The key configuration parameters are summarized in Table.~\ref{tab:env_setting}, which define the road structure, background vehicle behavior, and task-specific constraints. 

\begin{table}[htp]
    \centering
    \input{tabs/env_setting}
    \caption{Environment Configurations used in MAPPO training.}
    \label{tab:env_setting}
\end{table}

\subsubsection{Training Hyperparameters}
Table~\ref{tab:rl_training_params} summarizes the hyperparameters used in our MAPPO training for adversarial vehicle agents. These configurations are kept consistent across all environments.
\begin{table}[htp]
    \centering
    \input{tabs/RL_train_details}
    \caption{Hyperparameters used in MAPPO training.}
    \label{tab:rl_training_params}
\end{table}

\begin{figure*}[ht]
  \centering
  \includegraphics[width=0.8\textwidth]{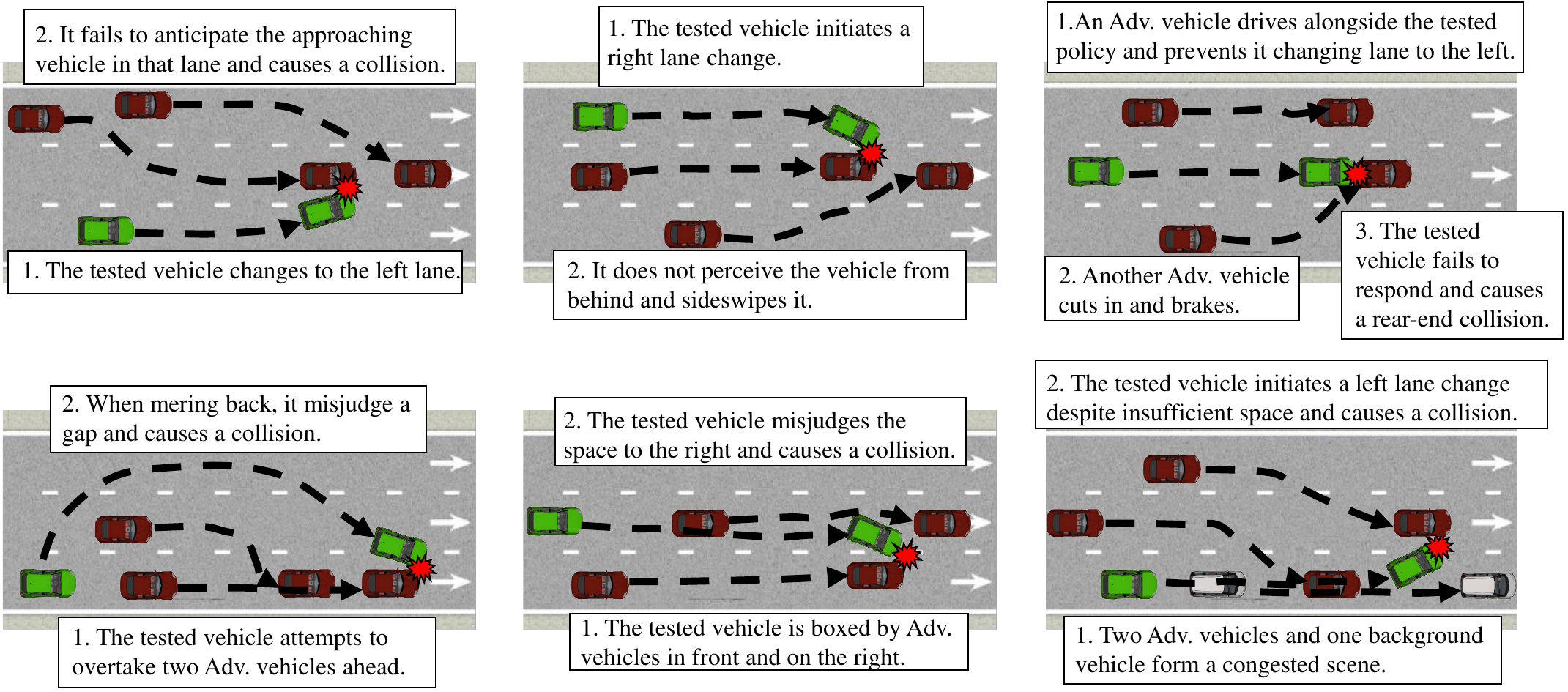}
  \caption{Examples of distinct vulnerability types discovered by AED. Adv. is abbreviation for             ``adversarial''.} 
  \label{fig:visualize_traj}
\end{figure*}

\begin{figure}[htp]
    \centering
    \includegraphics[width=0.25\textwidth]{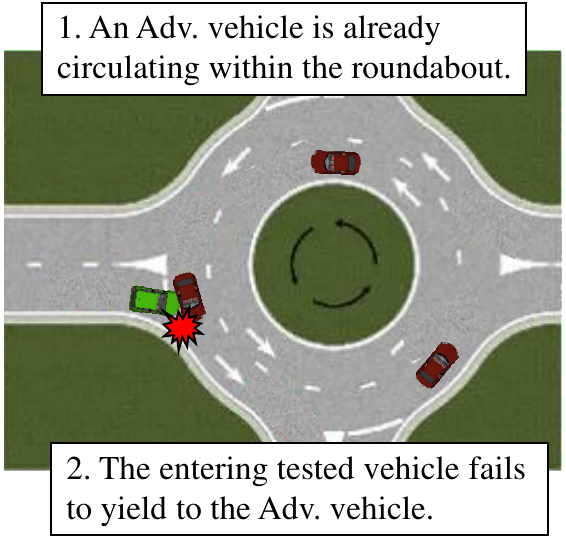}
    \caption{Example of T-Bone collision.}
    \label{fig:t_bone_visualize}
\end{figure}

\subsection*{Illustrations of Discovered Vulnerability Scenarios}

To better reflect real-world traffic incident patterns, we further decompose the three high-level categories---Left Lane Change, Right Lane Change, and Rear End collisions---into six fine-grained subtypes. These subtypes enable a more detailed assessment of vulnerability diversity:
\begin{itemize}
    \item Left-Sparse: The tested vehicle performs a left lane change without properly estimating the distance of an approaching adversarial vehicle in the adjacent lane, resulting in a sideswipe collision in sparse traffic.
    \item Right-Sparse: The tested policy executes a right lane change while misjudging the lateral distance to a nearby adjacent vehicle, causing under sparse traffic conditions.
    \item Rear End: The tested vehicle fails to maintain a safe following distance or reacts too slowly, resulting in a rear-end collision with an adversarial vehicle in the same lane.
    \item Overtake: The tested policy attempts to overtake a slower adversarial vehicle ahead, but misjudges the timing and crashes during the maneuver, typically due to its failure to account for adversarial vehicles blocking the return lane.
    \item Left-Dense: When facing congestion caused by an adversarial vehicle ahead, the tested policy attempts a left lane change but collides with another adversarial vehicle in the adjacent lane.
    \item Right-Dense: Similarly, due to congestion caused by a leading adversarial vehicle, the tested policy attempts a right lane change and causes a sideswipe collision.
\end{itemize}
Fig.~\ref{fig:visualize_traj} shows example trajectories for each of the above vulnerability subtypes in the Highway Environment. In the Roundabout environment, overtaking accidents are really rare in real-world scenarios and are therefore excluded from our analysis. Instead, we identify a distinct type of vulnerability unique to this setting as shown in Fig.~\ref{fig:t_bone_visualize}:
\begin{itemize}
    \item T-Bone: The tested vehicle misjudges the right-of-way when entering the roundabout, and fails to yield to an adversarial vehicle already circulating within it. This results in a perpendicular collision where the front of the tested vehicle crashes into the side of another.
\end{itemize}

\begin{table*}[htp]
    \centering
    \scriptsize
    \input{tabs/effective_appen}
    \caption{Full breakdown of effective Vulnerability Rates (standard deviation) for all fine-grained subtypes. AED consistently achieves the highest effective vulnerability rates.}
    \label{tab:effective}
\end{table*}

\subsection*{Full Breakdown of Effective Vulnerability Rates}
To complement the main text, we provide detailed results for all fine-grained vulnerability subtypes across environments, tested policies, and adversarial vehicles. The full breakdown, as presented in Table~\ref{tab:effective}, highlights that AED outperforms or matches other baselines across nearly all individual subtypes.




\subsection*{Prompt for LLMs}

\label{appendix: LLM prompts}

The prompts used to guide the LLM in generating diverse collision scenarios are shown in Prompt. \ref{prompt: 1} and Prompt. \ref{prompt: 2}. 

After multiple queries to the LLM and incorporating traffic rules, we selected seven representative collision types that are both realistic and commonly encountered. These seven types constitute the core collision scenarios studied in this paper. The LLM was then prompted again to generate the final task descriptions for each scenario.

The prompts used to generating reward function for a specific environment and collision are presented in Prompt.\ref{prompt: 3}, \ref{prompt: 4} and \ref{prompt: 5}. The specific environment descriptions and task descriptions for all collision types are presented in Sec.\ref{appendix: LLM Description}.

~\\

\input{tabs/prompt}

\subsection*{Detailed Description}

\label{appendix: LLM Description}

The environment descriptions for the Highway and Roundabout settings are provided in Prompt.~\ref{description: 1} and~\ref{description: 2}, respectively. As an illustrative example, the detailed task description for a rear-end collision scenario is presented in Prompt.~\ref{description: 3}. The descriptions of other scenarios differ only in the text specifying the collision configurations, which are provided in Prompt.~\ref{description: 4}.

~\\

\input{tabs/description}


%% file: tabs/obs_space.tex
\begin{tabular}{ccc}
\toprule
\textbf{Feature} & \textbf{Description}  & \textbf{Range}\\
\midrule
$p$   & Binary indicator of vehicle presence  & $\{0, 1\}$                   \\
$x$   & Relative longitudinal position                     & $[-150, 150] \mathrm{m}$  \\  
$y$   & Relative lateral position                     & $[-16, 16]\mathrm{m}$     \\
$v_x$ & Relative velocity along x-axis          & $[-60, 60]\mathrm{m/s}$   \\
$v_y$ & Relative velocity along y-axis          & $[-60, 60]\mathrm{m/s}$   \\

\bottomrule
\end{tabular}

%% file: tabs/action_space.tex
\begin{tabular}{cc}
\toprule
\textbf{Index} & \textbf{Discription} \\
\midrule
0        & Change to the left lane \\
1        & Maintain current lane and speed \\
2        & Change to the right lane \\
3        & Accelerate \\
4        & Decelerate \\
\bottomrule
\end{tabular}

%% file: tabs/env_setting.tex
\begin{tabular}{cc}
\toprule
\textbf{Parameter} & \textbf{Value} \\
\midrule
Number of Lanes                      & 4 \\
Number of Background Vehicles        & 5 \\
Background Vehicle Control Policy    & Value Iteration \\
Simulation Frequency                 & 5$\mathrm{Hz}$\\
Max Task Length                      & 20 \\
\bottomrule
\end{tabular}

%% file: tabs/RL_train_details.tex
\begin{tabular}{cc}
\toprule
\textbf{Hyperparameter} & \textbf{Value} \\
\midrule
Network Architecture        & MLP+RNN \\
Hidden Layer Size           & 64 \\
Initial Learning Rate       & 5e-4 \\
Discount Factor             & 0.99 \\
GAE Parameter               & 0.95 \\
Gradient clipping           & 0.5 \\
Value Loss Coefficient      & 0.5 \\
Entropy Coefficient         & 0.01 \\
PPO Clipping Parameter      & 0.2 \\
PPO Epochs                  & 15 \\
Training Steps              & 10M \\
Rollout Threads             & 100 \\
Max Episode Length          & 40 \\
\bottomrule
\end{tabular}

%% file: tabs/effective_appen.tex
\begin{tabular}{ccccccccc}
\toprule
                            &                        &        & \multicolumn{3}{c}{2 Adversaries} & \multicolumn{3}{c}{3 Adversaries} \\
                            &                        &        & VI       & RVI       & D3QN      & VI       & RVI       & D3QN      \\
\midrule
\multirow{18}{*}{Highway}    & \multirow{3}{*}{Left-Sparse}  & VDARS  &    $34.24\%_{(0.09)}$      &    $61.88\%_{(0.52)}$       &    $25.81\%_{(0.07)}$      &      $44.38\%_{(0.14)}$    &   $79.51\%_{(0.10)}$        &       $58.24\%_{(0.18)}$   \\
                            &                        & Eureka 
                            &    $97.22\%_{(0.08)}$      
                            &    $100\%_{(0.00)}$     
                            &    $73.89\%_{(0.55)}$    
                            &    $97.22\%_{(0.08)}$     
                            &    $97.78\%_{(0.08)}$   
                            &    $32.22\%_{(0.57)}$       \\
                            &                        & \cellcolor{lightpink} AED    &   \cellcolor{lightpink} $100\%_{(0.00)}$     &     \cellcolor{lightpink} $100\%_{(0.00)}$      &     \cellcolor{lightpink} $90.00\%_{(0.04)}$     &       \cellcolor{lightpink} $100\%_{(0.00)}$   &        \cellcolor{lightpink} $100\%_{(0.00)}$   &     \cellcolor{lightpink} $88.89\%_{(0.05)}$     \\ \cline{2-9} \addlinespace[0.5ex]
                            & \multirow{3}{*}{Right-Sparse} & VDARS  &     $39.56\%_{(0.11)}$     &      $11.18\%_{(0.09)}$     &     $0.00\%_{(0.00)}$     &     $26.92\%_{(0.08)}$     &       $5.13\%_{(0.01)}$    &      $0.00\%_{(0.00)}$    \\
                            &                        & Eureka 
                            &    $90.00\%_{(0.00)}$      
                            &    $83.89\%_{(0.28)}$     
                            &    $90.56\%_{(0.39)}$    
                            &    \cellcolor{lightpink} $92.78\%_{(0.48)}$     
                            &    $75.56\%_{(0.31)}$   
                            &    $11.11\%_{(0.64)}$           \\
                            &                        & \cellcolor{lightpink} AED    &      \cellcolor{lightpink} $95.00\%_{(0.04)}$    &      \cellcolor{lightpink} $87.77\%_{(0.03)}$     &     \cellcolor{lightpink} $100\%_{(0.00)}$     &    \cellcolor{lightpink} $90.00\%_{(0.01)}$      &     \cellcolor{lightpink} $90.56\%_{(0.10)}$      &     \cellcolor{lightpink} $30.56\%_{(0.20)}$     \\ \cline{2-9} \addlinespace[0.5ex]
                            & \multirow{3}{*}{Rear}  & VDARS  &     $2.28\%_{(0.03)}$     &   $1.49\%_{(0.01)}$        &     $1.94\%_{(0.01)}$     &      $1.45\%_{(0.01)}$    &      $0.85\%_{(0.01)}$     &    $0.00\%_{(0.00)}$      \\
                            &                        & Eureka 
                            &    $53.33\%_{(0.72)}$      
                            &    $58.33\%_{(0.36)}$     
                            &    $13.89\%_{(0.15)}$    
                            &    $90.56\%_{(0.44)}$     
                            &    $47.22\%_{(0.39)}$   
                            &    $36.11\%_{(0.28)}$            \\
                            &                        & \cellcolor{lightpink} AED    &    \cellcolor{lightpink} $60.56\%_{(0.05)}$      &      \cellcolor{lightpink} $63.89\%_{(0.06)}$     &     \cellcolor{lightpink} $31.11\%_{(0.06)}$     &      \cellcolor{lightpink} $95.60\%_{(0.01)}$    &       \cellcolor{lightpink} $82.22\%_{(0.10)}$    &     \cellcolor{lightpink} $55.00\%_{(0.23)}$     \\ \cline{2-9} \addlinespace[0.5ex]
                            & \multirow{3}{*}{Overtake}  & VDARS  &     $0.00\%_{(0.00)}$     &   $0.00\%_{(0.00)}$        &     $0.00\%_{(0.00)}$     &      $0.00\%_{(0.00)}$    &      $0.00\%_{(0.00)}$     &    $0.00\%_{(0.00)}$      \\
                            &                        & Eureka 
                            &   $46.67\%_{(0.28)}$       
                            &   $48.33\%_{(0.57)}$
                            &   \cellcolor{lightpink}
                            $1.67\%_{(0.01)}$
                            &   $51.67\%_{(0.15)}$
                            &   $51.67\%_{(0.07)}$
                            &   $4.00\%_{(0.03)}$             \\
                            &                        & \cellcolor{lightpink} AED    &    \cellcolor{lightpink} $56.11\%_{(0.01)}$      &      \cellcolor{lightpink} $51.11\%_{(0.05)}$     & 
                            \cellcolor{lightpink} $1.00\%_{(0.01)}$     &      \cellcolor{lightpink} $56.00\%_{(0.07)}$    &       \cellcolor{lightpink} $58.89\%_{(0.08)}$    &     \cellcolor{lightpink} $6.00\%_{(0.02)}$     \\ \cline{2-9} \addlinespace[0.5ex]
                            & \multirow{3}{*}{Left-Dense}  & VDARS  &     $0.00\%_{(0.00)}$     &   $0.00\%_{(0.00)}$        &     $0.00\%_{(0.00)}$     &      $0.00\%_{(0.00)}$    &      $0.00\%_{(0.00)}$     &    $0.00\%_{(0.00)}$      \\
                            &                        & Eureka 
                            &    $35.00\%_{(0.28)}$     
                            &    \cellcolor{lightpink}
                            $40.00\%_{(0.36)}$
                            &    \cellcolor{lightpink}
                            $8.33\%_{(0.15)}$
                            &    $58.33\%_{(0.23)}$
                            &    $45.00\%_{(0.86)}$
                            &    \cellcolor{lightpink} $60.00\%_{(0.15)}$           \\
                            &                        & \cellcolor{lightpink} AED    &    \cellcolor{lightpink} $40.56\%_{(0.07)}$      &   $36.11\%_{(0.02)}$     &     \cellcolor{lightpink} $8.00\%_{(0.02)}$     &      \cellcolor{lightpink} $88.33\%_{(0.03)}$    &       \cellcolor{lightpink} $78.89\%_{(0.05)}$    &      $47.77\%_{(0.07)}$     \\ \cline{2-9} \addlinespace[0.5ex]
                            & \multirow{3}{*}{Right-Dense}  & VDARS  &     $0.00\%_{(0.00)}$     &   $0.00\%_{(0.00)}$        &     $0.00\%_{(0.00)}$     &      $0.00\%_{(0.00)}$    &      $0.00\%_{(0.00)}$     &    $0.00\%_{(0.00)}$      \\
                            &                        & Eureka 
                            &   $70.00\%_{(0.23)}$      
                            &   \cellcolor{lightpink}
                            $75.00\%_{(0.05)}$
                            &   $20.00\%_{(0.07)}$
                            &   $28.33\%_{(0.23)}$
                            &   $75.00\%_{(0.34)}$
                            &   $30.00\%_{(0.54)}$            \\
                            &                        & \cellcolor{lightpink} AED    &    \cellcolor{lightpink} $76.11\%_{(0.03)}$      &      \cellcolor{lightpink} $75.00\%_{(0.00)}$     &     \cellcolor{lightpink} $35.56\%_{(0.08)}$     &      \cellcolor{lightpink} $61.11\%_{(0.43)}$    &       \cellcolor{lightpink} $82.22\%_{(0.05)}$    &     \cellcolor{lightpink} $30.56\%_{(0.20)}$     \\
\midrule
\multirow{18}{*}{Roundabout} & \multirow{3}{*}{Left-Sparse}  & VDARS  &       $2.57\%_{(0.02)}$   &     $1.67\%_{(0.01)}$      &   $0.58\%_{(0.01)}$       &    $1.54\%_{(0.01)}$      &      $2.70\%_{(0.03)}$     &     $0.00\%_{(0.00)}$     \\
                            &                        & Eureka 
                            &    \cellcolor{lightpink} $7.22\%_{(0.08)}$      
                            &    $5.00\%_{(0.08)}$     
                            &    $8.89\%_{(0.20)}$    
                            &    $1.11\%_{(0.15)}$     
                            &    $1.11\%_{(0.07)}$   
                            &    $0.00\%_{(0.00)}$       \\
                            &                        & \cellcolor{lightpink} AED    &        \cellcolor{lightpink} $6.67\%_{(0.01)}$  &     \cellcolor{lightpink} $8.33\%_{(0.03)}$      &     \cellcolor{lightpink} $8.89\%_{(0.01)}$     &     \cellcolor{lightpink} $3.33\%_{(0.01)}$     &    \cellcolor{lightpink} $4.44\%_{(0.01)}$       &      $0.00\%_{(0.00)}$    \\ \cline{2-9} \addlinespace[0.5ex]
                            & \multirow{3}{*}{Right-Sparse} & VDARS  &     $11.07\%_{(0.07)}$     &    $12.23\%_{(0.07)}$       &   $1.46\%_{(0.03)}$       &    $13.31\%_{(0.02)}$      &       \cellcolor{lightpink} $17.01\%_{(0.19)}$    &   $0.00\%_{(0.00)}$       \\
                            &                        & Eureka 
                            &    \cellcolor{lightpink} $12.78\%_{(0.28)}$      
                            &    $8.33\%_{(0.13)}$     
                            &    $2.22\%_{(0.08)}$    
                            &    $17.22\%_{(0.61)}$     
                            &    $11.11\%_{(0.15)}$   
                            &    $5.00\%_{(0.00)}$           \\
                            &                        & \cellcolor{lightpink} AED    &        $11.67\%_{(0.05)}$  &    \cellcolor{lightpink} $10.56\%_{(0.01)}$       &   \cellcolor{lightpink} $3.89\%_{(0.01)}$       &     \cellcolor{lightpink} $24.44\%_{(0.04)}$     &     \cellcolor{lightpink} $16.11\%_{(0.03)}$      &     \cellcolor{lightpink} $8.33\%_{(0.01)}$     \\ \cline{2-9} \addlinespace[0.5ex]
                            & \multirow{3}{*}{Rear}  & VDARS  &      $6.13\%_{(0.04)}$    &    \cellcolor{lightpink} $13.90\%_{(0.08)}$       &    $5.26\%_{(0.09)}$      &    $10.74\%_{(0.01)}$      &        $7.29\%_{(0.08)}$   &     $0.00\%_{(0.00)}$     \\
                            &                        & Eureka 
                            &    $3.33\%_{(0.36)}$      
                            &    $2.78\%_{(0.08)}$     
                            &    $15.56\%_{(0.43)}$    
                            &    $5.56\%_{(0.20)}$     
                            &    $7.78\%_{(0.20)}$   
                            &    \cellcolor{lightpink} $5.56\%_{(0.40)}$           \\
                            &                        & \cellcolor{lightpink} AED    &    \cellcolor{lightpink} $6.67\%_{(0.01)}$      &     $3.33\%_{(0.03)}$      &    \cellcolor{lightpink} $21.67\%_{(0.02)}$      &   \cellcolor{lightpink} $17.78\%_{(0.05)}$       &    \cellcolor{lightpink} $11.67\%_{(0.04)}$       &    \cellcolor{lightpink} $5.00\%_{(0.01)}$      \\ \cline{2-9} \addlinespace[0.5ex]
                            & \multirow{3}{*}{T-Bone}  & VDARS  &     $0.00\%_{(0.00)}$     &   $0.00\%_{(0.00)}$        &     $0.00\%_{(0.00)}$     &      $0.00\%_{(0.00)}$    &      $0.00\%_{(0.00)}$     &    $0.00\%_{(0.00)}$      \\
                            &                        & Eureka 
                            &    $8.00\%_{(0.04)}$      
                            &    $3.33\%_{(0.02)}$     
                            &    $2.00\%_{(0.08)}$    
                            &          \cellcolor{lightpink}
                            $3.33\%_{(0.15)}$     
                            &    $6.67\%_{(0.04)}$   
                            &    $0.00\%_{(0.00)}$            \\
                            &                        & \cellcolor{lightpink} AED    &    \cellcolor{lightpink} $11.67\%_{(0.05)}$      &      \cellcolor{lightpink} $8.33\%_{(0.063)}$     &     \cellcolor{lightpink} $3.33\%_{(0.01)}$     &      \cellcolor{lightpink} $2.78\%_{(0.01)}$    &       \cellcolor{lightpink} $10.00\%_{(0.05)}$    &  $0.00\%_{(0.00)}$     \\ \cline{2-9} \addlinespace[0.5ex]
                            & \multirow{3}{*}{Left-Dense}  & VDARS  &     $0.00\%_{(0.00)}$     &   $0.00\%_{(0.00)}$        &     $0.00\%_{(0.00)}$     &      $0.00\%_{(0.00)}$    &      $0.00\%_{(0.00)}$     &    $0.00\%_{(0.00)}$      \\
                            &                        & Eureka 
                            &  $0.00\%_{(0.00)}$      
                            &  $0.00\%_{(0.00)}$
                            &  $0.00\%_{(0.00)}$
                            &  $1.00\%_{(0.02)}$      
                            &  \cellcolor{lightpink}
                            $1.67\%_{(0.05)}$
                            &  $0.00\%_{(0.00)}$             \\
                            &                        & \cellcolor{lightpink} AED    &    $0.00\%_{(0.00)}$      &    $0.00\%_{(0.00)}$     &      $0.00\%_{(0.00)}$     &      \cellcolor{lightpink} $3.33\%_{(0.01)}$    &       \cellcolor{lightpink} $1.67\%_{(0.01)}$    &     $0.00\%_{(0.00)}$     \\ \cline{2-9} \addlinespace[0.5ex]
                            & \multirow{3}{*}{Right-Dense}  & VDARS  &     $0.00\%_{(0.00)}$     &   $0.00\%_{(0.00)}$        &     $0.00\%_{(0.00)}$     &      $0.00\%_{(0.00)}$    &      $0.00\%_{(0.00)}$     &    $0.00\%_{(0.00)}$      \\
                            &                        & Eureka 
                            &  $0.00\%_{(0.00)}$      
                            &  $0.00\%_{(0.00)}$
                            &  $0.00\%_{(0.00)}$
                            &  \cellcolor{lightpink}
                            $1.67\%_{(0.03)}$      
                            &  $1.11\%_{(0.01)}$
                            &  $1.00\%_{(0.02)}$     \\
                            &                        & \cellcolor{lightpink} AED    &    $0.00\%_{(0.00)}$      &    $0.00\%_{(0.00)}$     &    $0.00\%_{(0.00)}$     &      \cellcolor{lightpink} $1.00\%_{(0.01)}$    &       \cellcolor{lightpink} $1.67\%_{(0.02)}$    &    \cellcolor{lightpink} $1.11\%_{(0.02)}$     \\ 
\bottomrule
\end{tabular}

%% file: tabs/prompt.tex
\definecolor{codegray}{rgb}{0.5,0.5,0.5}
\definecolor{backcolour}{RGB}{245,245,245}

\lstdefinestyle{mystyle}{
    backgroundcolor=\color{backcolour},   
    commentstyle=\color{magenta},
    keywordstyle=\color{blue},
    numberstyle=\tiny\color{codegray},
    numbers=none,
    basicstyle=\fontfamily{\ttdefault}
    \footnotesize,
    breakatwhitespace=false,         
    breaklines=true,                 
    keepspaces=true,    
    frame=single,
    numbersep=5pt,                  
    showspaces=false,                
    showstringspaces=false,
    showtabs=false,                  
    tabsize=2,
    classoffset=1, %
    keywordstyle=\color{violet},
    classoffset=0,
}
\lstset{style=mystyle}
\renewcommand\lstlistingname{Prompt}

\begin{lstlisting}[
  basicstyle=\fontfamily{\ttdefault}\scriptsize,
  breaklines=true,
  caption={System Prompts of Generating Diverse Collision Patterns},
  label=prompt: 1
]
You are a helpful assistant designed to generate diverse collision scenarios for an autonomous driving simulation involving a defender, attackers, and dummy vehicles.
Your task is to generate different types of collisions based on traffic regulations where the defender is at fault. 
These collision scenarios will be used to write a reward function for reinforcement learning to improve the defender's strategy. 
You should consider various collision types, ensuring that the defender is primarily responsible according to traffic rules.
\end{lstlisting}

\begin{lstlisting}[
  basicstyle=\fontfamily{\ttdefault}\scriptsize,
  breaklines=true,
  caption={Prompts of Generating Diverse Collision Patterns},
  label=prompt: 2
]
I am working on a simulation environment that models autonomous driving strategies. The description of the environment is shown below.

<Environment Description>

I hope to write a reward function for the attackers to facilitate reinforcement learning for the following goal. 
My goal is to generate a variety of collision scenarios as many diverse collision scenarios between the defender and attackers during the simulation. 
However, the key requirement is that, according to traffic regulations, the defender should bear the primary responsibility for these collisions. In other words, the defender must be at fault under traffic rules, which will help identify vulnerabilities in the defender's strategy for further improvement.
Therefore, the task here is to propose different types of collisions that could occur in this environment. The collisions should be varied in nature, easy to determine based on vehicle states (e.g., position, speed, lane), and focused on the defender being primarily at fault. 
Examples of collisions could involve blocking an attacker's lane, sudden speed changes that force collisions, or any other traffic violation where the defender is responsible. 
I need these types for use in writing a reward function for reinforcement learning and for simulating realistic traffic behavior. 
Please sort all collision types by their likelihood of occurrence, from highest to lowest, and output the result.
Additionally, please format your response in a structured way, such that it can be easily extracted and used programmatically. I would like the response to follow this structure:

```json
{
  "collision_types": [
    {
      "type": "Collision Type Name",
      "description": "A detailed description of the collision type.",
      "fault": "Who is at fault for this collision (typically the defender).",
      "scenario": "A sample scenario where this collision would occur. It should contain specific actions of the attackers and the defender before and during the collision."
      "criteria": "Explain how to determine whether a collision belongs to the current type."
    },
    {
      "type": "Another Collision Type Name",
      "description": "Another description.",
      "fault": "Who is at fault.",
      "scenario": "Sample scenario."
      "criteria": "Explain how to determine whether a collision belongs to the current type."
    }
  ]
}
```

Here is a possible example for lane change collisions:
{
  "collision_types": [
    "type":  "Lane Change Collisions", 
    "description": "The defender changes lanes unexpectedly, causing a collision with an attacker or another vehicle already occupying that lane."
    "fault": "The defender is primarily at fault, as it violated traffic rules by changing lanes unsafely."
    "scenario": "The defender makes a lane change to the left or right sufficient clearance, directly causing the attacker or a dummy vehicle to crash into the defender."
    "criteria": "Before the collision's timestep, the defender and attacker should not be in the same lane; at the collision's timestep, the defender and attacker should be in the same lane; at the collision's timestep, the attacker should not have accelerated, decelerated, or changed lanes."
  ]
}
\end{lstlisting}

\begin{lstlisting}[
  basicstyle=\fontfamily{\ttdefault}\scriptsize,
  breaklines=true,
  caption={System Prompts of Generating Reward Function},
  label=prompt: 3
]
You are a reward engineer trying to write reward functions to solve reinforcement learning tasks as effective as possible.
Your goal is to write a reward function for the environment that will help the agent learn the task described in text. 
When provided with class methods or incomplete code, you are required to complete the code to the best of your understanding based on common patterns and conventions.
Your reward function should use useful variables and class functions from the environment by access the local variables of the class through `controlled_vehicles` for example.
The reward function signature can be: 

def compute_reward(self) -> Tuple[torch.Tensor, Dict[str, torch.Tensor]]:
    ...
    return reward, {}

Make sure any new tensor or variable you introduce is on the same device as "self.speed".
\end{lstlisting}

\begin{lstlisting}[
  basicstyle=\fontfamily{\ttdefault}\scriptsize,
  breaklines=true,
  caption={Prompts of Generating Reward Function},
  label=prompt: 4
]
The environment description is as follows:

<Environment Description> 

and the specific python environment code is 

<Environment Python Code>

Write a reward function for the following task: 

<Task Description>
\end{lstlisting}

\begin{lstlisting}[
  basicstyle=\fontfamily{\ttdefault}\scriptsize,
  breaklines=true,
  caption={Prompts of Tips for Writing Reward Function},
  label=prompt: 5
]
The output of the reward function should consist of two items:
    (1) the total reward (should be a tensor of shape (num_attackers, 1) ),
    (2) a dictionary of each individual reward component.
The code output should be formatted as a python code string: "```python ... ```".
Some helpful tips for writing the reward function code:
    (1) You may find it helpful to normalize the reward to a fixed range by applying transformations like torch.exp to the overall reward or its components
    (2) torch.norm can only be used on a tensor. Therefore, if the input is of type numpy.ndarray, you need to first convert it to a tensor, pass it to torch.norm, and then convert it back to ndarray (if needed).
    (3) If you choose to transform a reward component, then you must also introduce a temperature parameter inside the transformation function; this parameter must be a named variable in the reward function and it must not be an input variable. Each transformed reward component should have its own temperature variable
    (4) Make sure the type of each input variable is correctly specified; a float input variable should not be specified as torch.Tensor
    (5) Most importantly, the reward code's input variables must contain only attributes of the provided environment class definition (namely, variables that have prefix self.). Under no circumstance can you introduce new input variables.
    (6) Your reward should contain a component which rewards the defender-responsible collision and you should carefully consider how to determine whether a collision is defender-responsible. If a defender-responsible collision is created, you must grant a reward of 100 to all attackers. 
    (7) Aside from this component for defender-responsible collision, you should normalize all other reward components to the range [0,10] using transformations such as torch.exp.
\end{lstlisting}

%% file: tabs/description.tex
\definecolor{codegray}{rgb}{0.5,0.5,0.5}
\definecolor{backcolour}{RGB}{245,245,245}

\lstdefinestyle{mystyle}{
    backgroundcolor=\color{backcolour},   
    commentstyle=\color{magenta},
    keywordstyle=\color{blue},
    numberstyle=\tiny\color{codegray},
    numbers=none,
    basicstyle=\fontfamily{\ttdefault}
    \footnotesize,
    breakatwhitespace=false,         
    breaklines=true,                 
    keepspaces=true,    
    frame=single,
    numbersep=5pt,                  
    showspaces=false,                
    showstringspaces=false,
    showtabs=false,                  
    tabsize=2,
    classoffset=1, %
    keywordstyle=\color{violet},
    classoffset=0,
}
\lstset{style=mystyle}
\renewcommand\lstlistingname{Prompt}

\begin{lstlisting}[
  basicstyle=\fontfamily{\ttdefault}\scriptsize,
  breaklines=true,
  caption={Environment Description of Highway},
  label=description: 1
]
This environment is a simulation environment designed to model autonomous driving strategies. 
  
This is a one-way, four-lane highway, with travel occurring in the positive x-axis direction. The right side of the travel direction corresponds to the positive y-axis direction. The width of each lane is 4 and the length of each vehicle is 5. 
  
Multiple vehicles operate within the lanes based on their respective autonomous driving strategies. Among these vehicles, there is 1 defender, {n_attackers} attackers, and the remaining are dummy vehicles.

At any given moment, a vehicle can perform one of the following actions:

'LANE_LEFT': Change lanes to the left,
'IDLE': Remain idle,
'LANE_RIGHT': Change lanes to the right,
'FASTER': Accelerate, OR
'SLOWER': Decelerate.
\end{lstlisting}

\vspace{1em}

\begin{lstlisting}[
  basicstyle=\fontfamily{\ttdefault}\scriptsize,
  breaklines=true,
  caption={Environment Description of Roundabout},
  label=description: 2
]
This environment is a simulation environment designed to model autonomous driving strategies. 
  
In this task, the ego-vehicle if approaching a roundabout with flowing traffic. It will follow its planned route automatically, but has to handle lane changes and longitudinal control to pass the roundabout as fast as possible while avoiding collisions. The width of each lane is 4 and the length of each vehicle is 5. 

Multiple vehicles operate within the lanes based on their respective autonomous driving strategies. Among these vehicles, there is 1 defender, {n_attackers} attackers, and the remaining are dummy vehicles.

At any given moment, a vehicle can perform one of the following actions:

'LANE_LEFT': Change lanes to the left,
'IDLE': Remain idle,
'LANE_RIGHT': Change lanes to the right,
'FASTER': Accelerate, OR
'SLOWER': Decelerate.
\end{lstlisting}

\vspace{1em}

\begin{lstlisting}[
  basicstyle=\fontfamily{\ttdefault}\scriptsize,
  breaklines=true,
  caption={Task Description of Rear End Collision},
  label=description: 3
]
Your task is to design a reward function for the attackers to facilitate reinforcement learning. 

The goal is to generate as many divers collision scenarios between the defender and attackers(collisions between the defender and other dummy vehicles are also ok) as possible during the simulation. 

However, the key requirement is that, according to traffic regulations, the defender should bear the primary responsibility for these collisions. In other words, the defender must be at fault under traffic rules, which will help identify vulnerabilities in the defender's strategy for further improvement.

Specifically, we aim to maxmize the follwing type of collision, its key information is as follows:

{
"type": "Rear End Collision",
"description": "The defender collides with the attacker from behind, typically caused by the defender not maintaining a safe following distance or being unable to stop in time.",
"fault": "The defender is primarily at fault, as it failed to maintain an adequate following distance or was unable to react in time to avoid the collision.",
"scenario": "The defender is traveling on the same straight line as the attacker but from behind (at a lower x-coordinate). The attacker remains in the same lane throughout the event.",
"criteria": "Before the collision's timestep, the defender and attacker should be on the same straight path (same lane), with the defender positioned behind the attacker (lower x-coordinate). At the collision's timestep, the defender should be in the same lane as the attacker, but with the defender's position behind. The attacker should not have performed any significant acceleration or lane change prior to the collision. The defender should not change lanes before the collision, and at the time of the collision, the defender's and attacker's y-coordinates should be close to each other."
}
  
When you try to determine responsibility for an accident, you can simply check whether it is collision at the above type, and then give the attacker a corresponding reward based on whether the defender is primarily responsible.
  
Here are some suggestions for designing the reward function:

1. Access the state of each vehicle.
  You can access the state of each vehicle at the current step by using self.controlled_vehicles. 
  This is a list of MDPVehicle objects, where self.controlled_vehicles[0] is the defender, self.controlled_vehicles[self.n_defenders:self.n_defenders + self.n_attackers] are attackers.

  You can see the definition of the Vehicle, ControlledVehicle(inherited from Vehicle) and MDPVehicle(inherited from ControlledVehicle) in the environtment code to get more information about the attributes of the vehicle and some useful class function you can use.

  Besides, you can access self.controlled_vehicles_trajectory[-1] to get the trajectory of all vehicles in the last step, which is also a list of MDPVehicle objects. 
  Then you can compare the state of the vehicles in the last step and the current step to determine the action of the vehicle in the last step.

2. Rewarding Attackers for Generating Defender-Responsible Collisions
  You should design a component of the reward function that rewards attackers for successfully creating collision scenarios where the defender is at fault. 
  To achieve this, you can access the state of each vehicles, analyze whether there is a collision between the defender and attackers and then determine whether the defender is responsible for the collision.
  This will be a challenging task, so you are encouraged to create a relatively complex constraint to assess whether the defender is responsible for the collision. If the collision scenario meets the criteria, reward the attackers generously. If not, reward them less. 
  To determine if the defender is at fault, you should base the judgment on traffic rules. If this proves difficult, consider whether the defender could have avoided the collision by adjusting their strategy beforehand. For example, if the defender makes an incorrect decision, such as an unnecessary lane change or failure to decelerate in time, resulting in a collision, then the defender should be deemed responsible.
  
3. Avoiding Unnecessary Collisions
  The reward function should avoid rewarding collisions that occurring between attackers or between attackers and dummy vehicles.

4. Encouraging Attackers to Stay Close to the Defender
  You should also design a component of the reward function that encourages attackers to remain close to the defender throughout the simulation, such as a distance-based reward. Without this, attackers may fail to generate any collisions and, as a result, may not receive any reward.

5. You are also encouraged to design additional components for the reward function to further enhance the learning process.
    
6. Since the accident scenarios are generated by the colloboration of the attackers, you should reward all the attackers for an defender-responsible accident scenario.

7. If the attacker performs some erroneous actions, such as changing lanes or slowing down, at the moment of collision with the defender, it is highly likely that the attacker should also bear some responsibility for the incident. You need to carefully assess this situation. If the attacker is determined to be at fault in the current accident, then the attacker should receive a smaller reward or no reward at all.
\end{lstlisting}

\begin{lstlisting}[
  basicstyle=\fontfamily{\ttdefault}\scriptsize,
  breaklines=true,
  caption={Configurations of All Collisions},
  label=description: 4
]
Left-Sparse:
{
    "type": "Left Lane Change Collisions",
    "description": "The defender unexpectedly changes lanes to the left(decreasing the lane index), causing a collision with an attacker or another vehicle already occupying that lane.",
    "fault": "The defender is primarily at fault, as they violated traffic rules by changing lanes unsafely to the left without proper clearance.",
    "scenario": "The defender moves from their original lane to the left (decreasing the lane index), directly causing the attacker or another vehicle to crash into the defender's vehicle due to insufficient clearance or the attacker's inability to avoid the collision.",
    "criteria": "Before the collision's timestep, the defender and attacker should not be in the same lane; at the collision's timestep, the defender and attacker should be in the same lane; at the collision's timestep, the attacker should not have accelerated, decelerated, or changed lanes."
}

Right-Sparse:
{
    "type": "Right Lane Change Collisions",
    "description": "The defender unexpectedly changes lanes to the right(increasing the lane index), causing a collision with an attacker or another vehicle already occupying that lane.",
    "fault": "The defender is primarily at fault, as they violated traffic rules by changing lanes unsafely to the right without proper clearance.",
    "scenario": "The defender moves from their original lane to the right (increasing the lane index), directly causing the attacker or another vehicle to crash into the defender's vehicle due to insufficient clearance or the attacker's inability to avoid the collision.",
    "criteria": "Before the collision's timestep, the defender and attacker should not be in the same lane; at the collision's timestep, the defender and attacker should be in the same lane; at the collision's timestep, the attacker should not have accelerated, decelerated, or changed lanes."
}

Rear End:
{
    "type": "Rear End Collision",
    "description": "The defender collides with the attacker from behind, typically caused by the defender not maintaining a safe following distance or being unable to stop in time.",
    "fault": "The defender is primarily at fault, as it failed to maintain an adequate following distance or was unable to react in time to avoid the collision.",
    "scenario": "The defender is traveling on the same straight line as the attacker but from behind (at a lower x-coordinate). The attacker remains in the same lane throughout the event.",
    "criteria": "Before the collision's timestep, the defender and attacker should be on the same straight path (same lane), with the defender positioned behind the attacker (lower x-coordinate). At the collision's timestep, the defender should be in the same lane as the attacker, but with the defender's position behind. The attacker should not have performed any significant acceleration or lane change prior to the collision. The defender should not change lanes before the collision, and at the time of the collision, the defender's and attacker's y-coordinates should be close to each other."
}

Overtake:
{
    "type": "Failed Overtake Return to Original Lane",
    "description": "The defender moves to an adjacent lane to overtake a vehicle, but fails to accelerate enough to pass safely and attempts to return to the original lane too soon, resulting in a side or rear-end collision with the original vehicle.",
    "fault": "The defender is at fault for not completing the overtaking maneuver properly and merging back into the original lane unsafely.",
    "scenario": "The defender is behind a slower attacker in Lane 2. It changes to Lane 3 (right) using 'LANE_RIGHT', intending to pass. However, the defender does not gain sufficient speed to get ahead. Despite that, it executes a 'LANE_LEFT' to return to Lane 2, where the attacker is still present. This causes a collision as the attacker impacts the side or rear of the defender.",
    "criteria": "Defender initially changes lanes to pass, then attempts to return to the original lane. The attacker or dummy vehicle remains in the original lane and does not change speed or lanes. The collision occurs during or immediately after the defender tries to return to the original lane. The x-position of the defender is still close to or behind the attacker. You can check self.controlled_vehicles_trajectory[-5:-1] to get the previous actions of the defender before the collision."
}

Left-Dense:
{
    "type": "Left Lane Change During Congestion",
    "description": "The defender attempts to switch to the left lane in dense traffic due to a slower vehicle ahead but fails to properly assess the space in the target lane. The defender merges into a gap that does not provide a safe buffer with the vehicle in the left lane, causing a side or rear-side collision.",
    "fault": "The defender is at fault for merging left without maintaining a safe gap in heavy traffic, violating lane-change safety protocols.",
    "scenario": "In a congested traffic setting, the defender's current lane has a slower vehicle directly ahead at a close distance. Seeking to maintain momentum, the defender initiates a 'LANE_LEFT' maneuver toward a seemingly faster adjacent lane. However, the left lane already contains a vehicle (attacker or dummy) very close to the intended merge location. The defender does not sufficiently reduce speed and clips the front or side of the vehicle in the left lane.",
    "criteria": "There must be high traffic density in both lanes (i.e., low spacing between vehicles). A vehicle must be present in close proximity ahead of the defender in its original lane. The defender must initiate a left lane change when there is insufficient gap to the vehicle ahead in the target lane. A collision must occur during or immediately after the lane change, with the point of impact on the defender's left rear-side or side, and the other vehicle's front or front-side."
  }

Right-Dense:
{
    "type": "Right Lane Change During Congestion",
    "description": "The defender attempts to switch to the right lane in dense traffic due to a slower vehicle ahead but fails to properly assess the space in the target lane. The defender merges into a gap that does not provide a safe buffer with the vehicle in the right lane, causing a side or rear-side collision.",
    "fault": "The defender is at fault for merging right without maintaining a safe gap in heavy traffic, violating lane-change safety protocols.",
    "scenario": "In a congested traffic setting, the defender's current lane has a slower vehicle directly ahead at a close distance. Seeking to improve progress, the defender initiates a 'LANE_RIGHT' maneuver toward a seemingly faster adjacent lane. However, the right lane already contains a vehicle (attacker or dummy) very close to the intended merge location. The defender does not sufficiently reduce speed and clips the front or side of the vehicle in the right lane.",
    "criteria": "There must be high traffic density in both lanes (i.e., low spacing between vehicles). A vehicle must be present in close proximity ahead of the defender in its original lane. The defender must initiate a right lane change when there is insufficient gap to the vehicle ahead in the target lane. A collision must occur during or immediately after the lane change, with the point of impact on the defender's right rear-side or side, and the other vehicle's front or front-side."
  }

T-Bone:
{
    "type": "T-Bone Collision at Intersection (Roundabout Only)",
    "description": "A side-impact (T-bone) collision occurs when the front of the defender hits the side of another vehicle (attacker or dummy) at an intersection or roundabout due to the defender ignoring right-of-way or entering improperly.",
    "fault": "The defender is at fault for not yielding at the intersection or roundabout, violating the right-of-way of the attacker or another vehicle already in the roundabout.",
    "scenario": "In a roundabout environment, the attacker is traveling along the roundabout lane, already engaged in the circular path. The defender enters the roundabout from an entry lane without yielding, performing a 'FASTER' or even an 'IDLE' to merge. The defender collides with the side of the attacker's vehicle at the intersection of their paths.",
    "criteria": "Defender should be at the roundabout entry point. Attacker is already in the roundabout, moving along its circular or curved path. Collision must involve the front of the defender and the side of the attacker. Attacker's speed and path should remain stable. Defender must enter without stopping or yielding."
  }
\end{lstlisting}